\documentclass[12pt,preprint]{aastex}

\begin{document}

\title{Non-Thermal Chemistry in Diffuse Clouds with Low Molecular 
Abundances$^1$}

\author{J. Zsarg\'{o}$^{2,\ 3}$ and S.R. Federman$^2$}

\begin{abstract}

High quality archival spectra of interstellar absorption from \ion{C}{1} 
toward 9 stars, taken with the Goddard High Resolution Spectrograph on the 
{\it Hubble Space Telescope}, were analyzed. Our sample was supplemented 
by two sight lines, 23~Ori and $\beta^1$~Sco, for which the \ion{C}{1} 
measurements of Federman, Welty, \& Cardelli were used. Directions 
with known CH$^+$ absorption, but only upper limits on absorption from 
C$_2$ and CN, were considered for our study. This restriction allows us 
to focus on regions where CH$^+$ chemistry dominates the production of 
carbon-bearing molecules.

Profile synthesis of several multiplets yielded column densities and 
Doppler parameters for the \ion{C}{1} fine structure levels. Equilibrium 
excitation analyses, using the measured column densities as well as the 
temperature from H$_2$ excitation, led to values for gas density. These 
densities, in conjunction with measurements of CH, CH$^+$, C$_2$, and 
CN column densities, provided estimates for the amount of CH associated 
with CH$^+$ production, which in turn set up constraints on the present 
theories for CH$^+$ formation in this environment. We found for our 
sample of interstellar clouds that on average, 30--40\% of the 
CH originates from CH$^+$ chemistry, and in some cases it can be 
as high as 90\%.

A simple chemical model for gas containing non-equilibrium production of 
CH$^+$ was developed for the purpose of predicting column densities for 
CH, CO, HCO$^+$, CH$_2^+$, and CH$_3^+$ generated from large abundances of 
CH$^+$. Again, our results suggest that non-thermal chemistry is necessary 
to account for the observed abundance of CH and probably that of CO in these 
clouds.

\end{abstract}


\small

$^1$\ Based on observations obtained with the NASA/ESA $Hubble$ 
$Space$ $Telescope$ through the Space Telescope Science Institute, which is 
operated by the Association of Universities for Research in Astronomy, Inc.,
under NASA contract NAS5-26555.

$^2$\ Department of Physics and Astronomy, University of Toledo, 
Toledo, OH 43606.

$^3$\ Department of Physics and Astronomy, The Johns Hopkins 
University, Baltimore, MD 21218.

\normalsize

\section{Introduction}

Carbon is rarely found in the form of \ion{C}{1} in the general interstellar 
medium (ISM). In the hot, ionized part of the ISM (like the interior of 
superbubbles), 
it is highly ionized, while in the warm neutral regions it is mainly in the 
form of C~{\small II}. As the cloud gets colder the recombination
of C~{\small II} and the transition from \ion{C}{1} to CO appears almost 
simultaneously (Hollenbach \& Tielens 1999), leaving little room for \ion{C}{1} 
as the dominant form of carbon. The diffuse
clouds where this double transition usually occurs are very important for 
understanding the nature of
atomic and molecular chemistry in the ISM. It is here where the important 
nonthermal effects, dissipation
of turbulence (Duley et al. 1992; Falgarone, Pineau des For\^{e}ts, 
\& Roueff 1995; Hogerheijde 
et al. 1995; Joulain et al. 1998; Pety \& Falgarone 2000) and 
MHD waves (e.g., Draine \& Katz 1986; Pineau des For\^{e}ts et al. 1986;
Gredel, van Dishoeck, \& Black 1993; Charnley \& Butner 1995; 
Federman et al. 1996a), can seriously alter the chemical composition 
of the clouds. For example, it is not possible to understand the observed 
abundance of CH$^+$ without taking into account these effects. 

Our first objective is to derive the gas density and the kinetic temperature 
for such diffuse clouds. We 
infer this information from the observed populations among fine structure 
levels of the electronic 
ground state of \ion{C}{1}. We searched the {\it Hubble Space Telescope} 
($HST$) archive for 
spectra of bright stars known to show weak or moderate molecular absorption to 
minimize the effects 
of denser molecular cores. In all, we found nine suitable target stars; their 
properties from the Simbad database are listed in Table~\ref{table:LOS}. 
The reddenings are from Savage et al. (1977). 
We included 23~Ori and $\beta^1$~Sco (not listed in 
Table~\ref{table:LOS})
in our sample with \ion{C}{1} column densities from Federman et al. (1997).

The target stars are located in Taurus, Perseus, Orion, Scorpius, and 
Ophiuchus, with the exception 
of HD~112244. The general environment toward these stars is well characterized 
through numerous studies. 
$\eta$~Tau, for example, is one of the brightest members of the Pleiades (White 
1984). Krelowski, Megier, \& 
Strobel (1996) found two absorbing components toward the Perseus OB2 
association, but simple molecules (CH, 
CN) could be detected only in one of them. 
UV observations of Martin \& York (1982) 
and Vidal-Madjar et al. (1982) detected several absorption components 
toward the Perseus association. 
With the exception of one cold component, these were too weak and too hot to be 
observable in neutral 
carbon. The temperatures deduced by Trapero et al. (1992) from interstellar 
K~{\small I} 
absorption toward stars in Perseus are in general agreement with the results of 
this study. 

de Geus (1992) described the immediate vicinity of the 
Scorpio-Centaurus region in detail. 
He found that the ISM has been shocked and disturbed by the strong winds of 
O and B stars and past 
Type~II supernovae from the numerous OB associations located here. 
A CO survey of the Ophiuchus region by de Geus, Bronfman, \& Thaddeus (1990) 
found 5 main and several weaker 
cloud components. It is likely, however, that some of them are dense molecular 
cores not relevant for 
this study. High resolution spectroscopy of Na~{\small I}~D and K~{\small I} 
$\lambda$7699 lines by Kemp, 
Bates, \& Lyons (1993) revealed only 3 components, two stronger and one weak, 
in general agreement with our findings of two components. 
The more recent atomic surveys of Welty, Hobbs, \& Kulkarni (1994) and 
Welty \& Hobbs (2000) also indicate that our C~{\small I} measurements 
probe the strongest components toward our sample of stars. 
$Copernicus$ observations of $\sigma$~Scorpii (Allen, Snow, 
\& Jenkins 1990) revealed 4 components based primarily on H$_2$ lines.   

The main objective of this study is to obtain information on the importance of 
non--equilibrium 
CH$^+$ chemistry for production of CH and CO in diffuse clouds. These 
molecules can be produced 
in very different environments. In the denser portions of a cloud they are 
involved in a complex 
thermal chemistry, while in warmer and more tenuous portions they are 
produced via CH$^+$. 
Since any line of sight can involve several clouds and cloud sections, the 
inferred quantities and 
parameters are likely the result of several environments. If one can set up 
reasonable estimates of 
the contributions from the different environments, further constraints on the 
physical processes underlying nonthermal CH$^+$ chemistry can be 
established. 

We follow two approaches in our study. On the one hand, we estimate the amount 
of CH produced in 
denser regions (with equilibrium chemistry) and establish lower and upper 
limits on CH produced by 
the molecular chemistry. The difference between our limits and the observations 
should come from 
non-equilibrium chemistry, and this has to be accommodated by any model 
attempting to explain
CH$^+$ chemistry.  On the other hand, we calculate CH and CO 
abundances arising from CH$^+$ in diffuse clouds 
using the model of Federman et al. (1996a) to have a rough estimate of the 
amount coming from non-equilibrium 
chemistry. As an improvement on the models in Federman et al. (1996a), 
we use increased 
effective temperatures not only for ions, but also for neutrals directly 
produced by ion--ion or ion--neutral reactions. The effects on CH were 
noted by Gredel et al. (1993) and described in detail by Bucher \& 
Glinski (1999).

Details of the data reduction and measurements are discussed in 
Section~\ref{section:Obs}. 
The gas densities and temperatures were inferred from relative populations 
obtained by fitting synthetic 
spectra to the observations. We discuss this analysis in 
Section~\ref{section:Model} and describe
how the gas densities and temperatures were inferred from \ion{C}{1} 
excitation in 
Section~\ref{section:Excitation}. We present the calculation of molecular 
abundances in 
Sections~\ref{section:EQC} and \ref{section:NEQC} for the equilibrium and 
non-equilibrium chemistry, respectively. We discuss our result briefly
in Section~\ref{section:Discussion} and summarize them in 
Section~\ref{section:Summary}. 

\section{Measurements and Data Reduction\label{section:Obs}}

We retrieved observations for 9 new lines of sight from the $HST$ archive; 
most were acquired at medium resolution (MR; R=$\frac{\lambda}{\Delta 
\lambda}$=20,000) and a handful
were taken at high resolution (HR; R= 100,000). 
These spectra covered several \ion{C}{1} multiplets and forbidden lines 
between 1150~\AA~ and 
1700~\AA. The datasets that were retrieved from the $HST$ archive and used in 
our study are presented in Table~\ref{table:Obs}. 

We used the standard IRAF STSDAS package to cross-correlate and coadd the 
subexposures of the archival spectra. 
The most serious concern 
was an error in the background correction on some of our preprocessed data. 
We detected this error on all 
of our HR spectra containing the $\lambda$1260 multiplet of 
\ion{C}{1}. The flux level 
of a strongly saturated line, such as those of S~{\small II} and 
Si~{\small II}, should be zero at line 
center; however, in some cases we measured negative or small positive 
values, clearly indicating an erroneous 
background correction. We could then apply an additional correction to 
compensate for the effect.  
We never encountered this problem on MR spectra although we had saturated 
lines in some cases, probably because 
scattered light was more easy to quantify in grating spectra.

After calibration and data reduction, the components in the neutral carbon 
lines were identified 
and the continuum of the background star was removed (continuum placement) 
by fitting it to a low 
order polynomial. The continuum placement turned out to be difficult for some 
multiplets, due to blending of several lines or simply because the lines 
were saturated. For example, in the 
vicinity of multiplet $\lambda$1260, Si~{\small II} and Fe~{\small II} 
lines seriously affected the continuum 
placement.

Following line identification and continuum placement, the Doppler shift 
(relative to the laboratory wavelength) 
and equivalent width ($W_{\lambda}$) of neutral carbon lines were measured. 
In some cases, only upper limits could be extracted from the average 
Full Width at Half Maximum (FWHM) of the instrumental function and the 
root mean square (rms) average of the 
noise in the stellar continuum ($W_{\lambda} \leq$ 
2~$\times$~FWHM~$\times$~rms). The measured values of $W_{\lambda}$
are listed in Tables~\ref{table:App1}-\ref{table:App2} in the Appendix 
for comparison with results of other investigations.

The $HST$ data for $\epsilon$ Per and $\omega^1$ Sco were supplemented 
with spectra for additional lines acquired with the $Copernicus$ satellite.  
These coadded spectra were obtained from the Multiwavelength Archive at 
the Space Telescope Science Institute (MAST).  All spectra for each of 
the two directions were analyzed together.

\section{\ion{C}{1} Column Densities and Doppler Parameters 
\label{section:Model}}

The column densities ($N$) and the Doppler parameters ($b$-values) that 
describe the absorbing components 
along each line of sight were determined by fitting synthetic profiles to 
the observed spectra. 
The synthetic fluxes were compared to the observations and the set of 
initial parameters was adjusted by a $\chi^2$ minimization until a 
satisfactory fit to all observed 
spectra for a sight line could be achieved. Our approach was a 
modified second order gradient method, 
taken from Bevington \& Robinson (1992). The program was tested by 
reproducing the results of curves of growth. 
We estimated the errors for the 
inferred parameters to be $\approx$ 
10\%, based on the goodness of the continuum placement and the 
noise in the spectra. The effects of other potential sources of error were 
negligible.  For oscillator strengths, we used those from the compilation of 
Morton (1991), updated with our more recent astronomical determinations 
(Zsarg\'{o}, Federman, \& Cardelli 1997; Federman \& Zsarg\'{o} 2001).

Throughout the synthesis we used a Gaussian instrumental function with a 
FWHM of $\approx$ 3 km s$^{-1}$ 
and $\approx$ 15 km s$^{-1}$ for HR and MR spectra, respectively. The exact 
values of FWHM were 
calculated for each spectrum from the wavelength dependence in the 
resolution of Echelle-A and 
G160M (described in the instrumental handbook for the GHRS).

\begin{figure}[htbp] 
\epsscale{0.75}\plotone{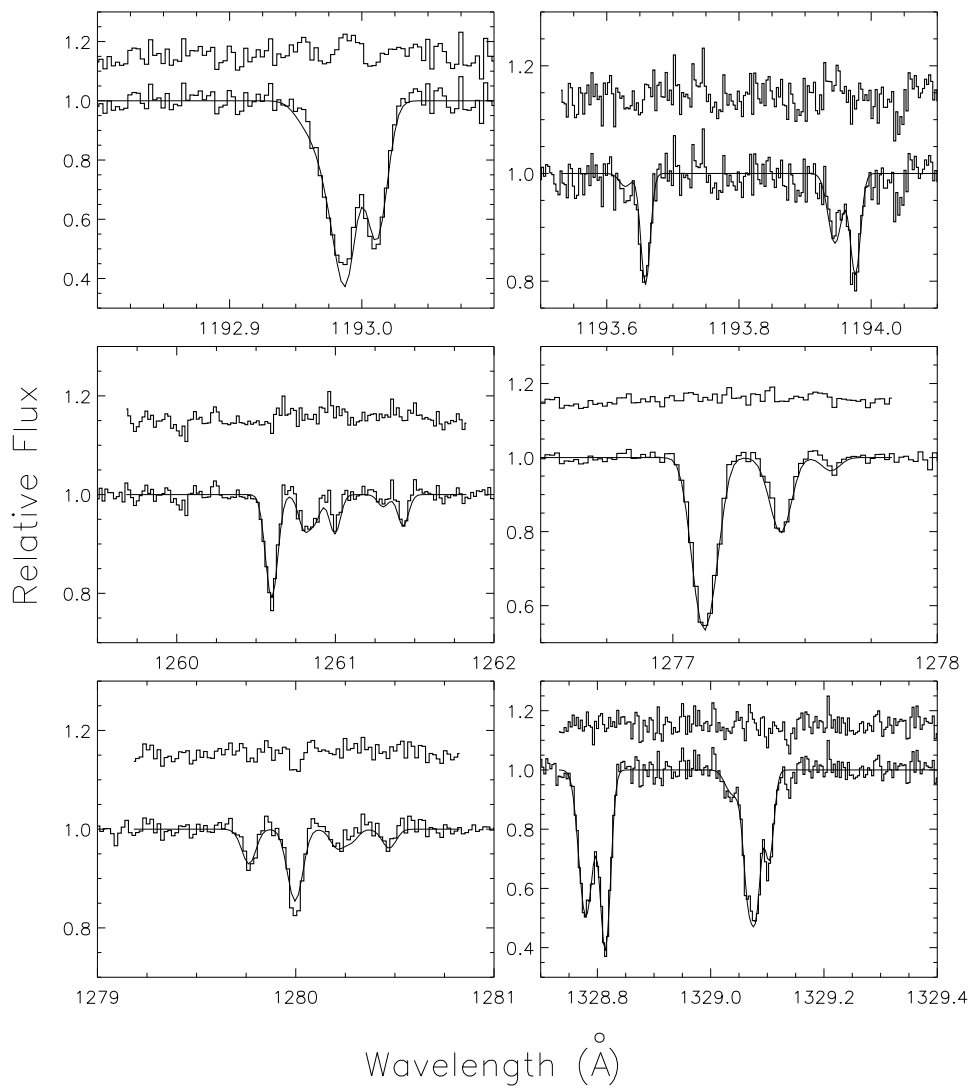}
\caption{Observed (histograms) and model (lines) spectra of 
the \ion{C}{1} multiplets toward 1~Sco. All spectra are medium 
resolution unless otherwise noted.
a) The top, middle, and bottom rows from left to right show multiplets
$\lambda$$\lambda$1193, 1194 (high resolution), 1260, 1277, 1280, and 1329 
(high resolution, $J$ $=$ 0 and 1 levels only), respectively. The residuals 
are displayed above each spectra, offset by +1.15. \label{fits} }
\end{figure}

The results of the profile syntheses are shown in 
Table~\ref{table:fit}. To demonstrate the quality of 
fits that could be achieved by this procedure, we display
observed and synthesized 
spectra for 1~Sco in Figure~\ref{fits}. The differences between the 
observed and model spectra in the Figure are well within the 
continuum noise for all \ion{C}{1} multiplets. The HR  spectra 
of $\lambda$~Ori, 1~Sco, $\delta$~Sco, and
$\sigma$~Sco revealed that there are at least two components in 
\ion{C}{1} absorption along these sight lines. The two components 
toward 1~Sco, for example, are clearly visible in the HR spectra of 
\ion{C}{1} $\lambda$1328.83 in Figure~\ref{fits}. We could not
discern multiple components toward most sight lines for which only MR
spectra were available. The only exception is HD~112244 
where the large separation ($\approx$~24 km s$^{-1}$)
between the components was easily recognizable. The main absorption 
component (Component A) for the lines of sight in Table~\ref{table:fit} 
are very well defined. The uncertainties in their columns
and $b$-values are dominated by the errors in the continuum placement and 
by the signal-to-noise ratio of the spectra. The parameter values for 
the secondary components (Component B), on the other 
hand, are not defined very well. 
Often, we had to keep some parameters fixed during the synthesis in order 
to achieve reasonable values for 
the column densities, or else, only upper limits of $N$ could be provided. 
 
Comparison between our line parameters and those extracted from high and 
ultra-high resolution measurements of atoms and molecules shows good 
correspondence.  There are four sight lines with one C~{\small I} component, 
$\eta$ Tau, $\epsilon$ Per, $\pi$ Sco, and $\omega^1$ Sco.  The sight lines 
toward $\eta$ Tau and $\epsilon$ Per indicate but one component in 
K~{\small I} (Welty \& Hobbs 2000) with $b$-values of about 1 km s$^{-1}$.  
Presumably, the larger $b$-values in C~{\small I} arise because this atom 
probes more widely distributed material.  For $\pi$ Sco and $\omega^1$ Sco, 
Welty et al. (1994) and Welty \& Hobbs (2000) see multiple components 
in Na~{\small I}~D and K~{\small I}, respectively.  Here, our lower 
resolution measurements could not distinguish absorption from the 
stronger, closely spaced components.  The $b$-values for the $J$ $=$ 0 
lines reflect this.  The narrower $b$-values for lines from $J$ $=$ 1 and 
2 are likely the result of their sampling only the strongest component, 
a smaller portion of the line of sight, or both.

A similar pattern emerges for the sight lines with two C~{\small I} 
components.  The $\approx$ 12 km s$^{-1}$ separation seen toward $\lambda$ 
Ori comes from the complexes discerned in Na~{\small I}~D measurements 
(Hobbs 1974; Welty \& Hobbs 2000).  Toward HD~112244, the two components 
about 24 km s$^{-1}$ apart are also present in CH spectra (Danks, 
Federman, \& Lambert 1984).  For 1 Sco, the correspondence is seen in 
Na~{\small I}~D complexes (Welty et al. 1994; Welty \& Hobbs 2000), 
separated by about 7 km s$^{-1}$.  The same applies to the two C~{\small I} 
components ($\Delta v \approx 5$ km s$^{-1}$) toward $\delta$ Sco, which 
are present in Na~{\small I}~D and K~{\small I} (Welty \& Hobbs 2000) and 
CH$^+$ (Crane, Lambert, \& Sheffer 1995).  Finally, the 9 km s$^{-1}$ 
separation in components toward $\sigma$ Sco is seen in 
Na~{\small I}~D and K~{\small I} (e.g., Welty \& Hobbs 2000).

\section{Excitation Analysis \label{section:Excitation}}

We used the column densities found from the profile syntheses to infer gas 
densities and temperatures for the gas containing \ion{C}{1}. 
To do so we assumed steady-state detailed balance 
among the fine structure levels $J$ $=$ 
0, 1, and 2 of neutral carbon. Two independent number density ratios can be
established. These are
\begin{equation}
\frac{n_{J=2}}{n_{J=0}}= \; \frac{  R_{0,2} \left( A_{1,0} + R_{1,0} + R_{1,2} 
\right) \; + \; R_{1,2} 
R_{0,1} }{ \left( A_{2,1} + R_{2,1} + R_{2,0} \right) \left( A_{1,0} 
+ R_{1,0} + R_{1,2} \right) \; - \; 
R_{1,2} \left( A_{2,1} + R_{2,1} \right)} \label{eq:ratio2}
\end{equation}
and
\begin{equation}
\frac{n_{J=1}}{n_{J=0}}= \; \frac{  R_{0,1} \left( A_{2,1} + R_{2,1} + R_{2,0} 
\right) \; + \; R_{0,2} 
\left( A_{2,1} + R_{2,1} \right) }{ \left( A_{2,1} + R_{2,1} + R_{2,0} \right) 
\left( A_{1,0} + R_{1,0} + 
R_{1,2} \right) \; - \; R_{1,2} \left( A_{2,1} + R_{2,1} \right)} \;\; , 
\label{eq:ratio1}
\end{equation}
where the terms $R_{i,j}$ are defined by
\begin{equation}
R_{i,j}= \sum_k \gamma^k_{i,j} n_k + G_{i,j} \;\; .
\end{equation}
The quantities $\gamma^k_{i,j}$ are the collisional excitation or 
de-excitation rate coefficients through 
collision with species k and $G_{i,j}$ are the ultraviolet radiative 
pumping rates between fine structure
levels i and j. In eqns.~\ref{eq:ratio2} and  \ref{eq:ratio1}, the quantity 
$A_{i,l}$ is the far 
infrared spontaneous decay rate from level i to level l. Variables $n_k$ 
are the number density of species k. For 
our lines of sight, collisions with H, He, and H$_2$ were considered, 
since they are the most abundant species in 
diffuse interstellar clouds. The \ion{C}{1} collisional de-excitation rate 
coefficients are listed in Table~\ref{table:collision}. The corresponding 
excitation rates can be calculated by the formula 

\begin{equation}
\gamma^k_{j,i} = \gamma^k_{i,j} \frac{2i+1}{2j+1} e^{-T_{i,j}/T} \;\; ,
\end{equation}
where $kT_{i,j}$ and $T$ are the energy between states i and j and the
gas temperature, respectively.
The rates for \ion{C}{1} collisions with H$_2$ and H were taken from 
Schr\"{o}der et al. (1991) and from 
Launay \& Roueff (1977), respectively. The 
collisional rates for He were from 
Lavendy, Robbe, \& Roueff (1991) and Staemmler \& Flower (1991). 
Radiative pumping was neglected in our
analysis because it has a negligible effect on the populations, 
unless the ultraviolet field is at 
least 10 times the strength of the average interstellar field. 

If the clouds have close to uniform density and chemical composition, then
\begin{equation}
n_k \approx \frac{N_k \, n}{N_H + N_{H_2} + N_{He}} \;\; , 
\label{eq:approximations}
\end{equation}

\begin{displaymath}
\frac{n_{J=1}}{n_{J=0}} \approx \frac{N_{J=1}}{N_{J=0}} \;\;\;\; and 
\;\;\;\; \frac{n_{J=2}}{n_{J=0}} \approx \frac{N_{J=2}}{N_{J=0}} \;\; ,
\end{displaymath} 
where the $N$-s are the column density of species k, hydrogen, H$_2$, 
helium, or a \ion{C}{1} 
fine structure level, and $n$ is the gas density. We calculated the fractional 
abundances for H and
H$_2$ using column densities from Savage et al. (1977) and we assumed that 
10\% of the material is He for all lines of sight. 

After substituting eqn.~\ref{eq:approximations} in 
eqns.~\ref{eq:ratio2} and \ref{eq:ratio1}, one has two equations 
for the unknown temperature and gas density. Unfortunately, 
the temperature and density solutions for eqns.~\ref{eq:ratio2} and 
\ref{eq:ratio1} are not unique. One can find many density-temperature
pairs to satisfy these equations for a specific $\frac{n_{J=1}}{n_{J=0}}$
or $\frac{n_{J=2}}{n_{J=0}}$ ratio. As a further complication, 
the uncertainties in the \ion{C}{1} column densities introduce 
uncertainties in the \ion{C}{1} level ratios; 
therefore, a density range rather than a given value satisfies 
eqn.~\ref{eq:ratio2} or \ref{eq:ratio1} for a temperature. As a
consequence of this ambiguity, numerous density-temperature pairs
simultaneously satisfy eqns.~\ref{eq:ratio2} and \ref{eq:ratio1}.
One needs independent density or temperature estimates to further
confine the solutions.
 
The density--temperature solutions of eqns.~\ref{eq:ratio2} and 
\ref{eq:ratio1} are displayed in Figure~\ref{DTcurves} for the main 
components of each sight line. The \ion{C}{1} excitation analysis was
supplemented by the temperature estimates of Savage et al. (1977) which
were determined by H$_2$ excitation analysis for the sight lines.
Since the H$_2$ temperatures of Savage et al. (1977) are valid only 
for the main components, we did not perform the \ion{C}{1} excitation 
analysis for the secondary components. 
The sole exception is HD~112244 where the two components are fairly 
similar; therefore, the temperature estimate of Savage et al. (1977) 
most likely applies to both components. In Figure~\ref{DTcurves},
each \ion{C}{1} ratio resulted in two density--temperature curves 
corresponding to the maximum and minimum ratios that were allowed 
by the uncertainties in the \ion{C}{1} column densities.
Any density-temperature point between the solid or the dashed curves 
in Figure~\ref{DTcurves} is consistent with the observed \ion{C}{1} 
$J$=0 and 1 or the $J$=0 and 2 column densities. 
Then, the region that is between all four curves contains all
density-temperature pairs that are the simultaneous solutions 
to eqns.~\ref{eq:ratio2} and \ref{eq:ratio1}. However, 
Figure~\ref{DTcurves} also shows that the temperature values are not 
well defined for most sight lines. The allowed temperatures
range between 30 and 200 K, and can be confined only by the 
estimates of Savage et al. (1977). 
For simplicity, we accepted as solutions all the density-temperature
pairs in Figure~\ref{DTcurves} that lie within the thick rectangles. 
These regions provide the smallest independent density and temperature 
ranges that satisfy both eqns.~\ref{eq:ratio2} and \ref{eq:ratio1} for the
observed \ion{C}{1} ratios and consistent with the H$_2$ excitation 
temperatures. The derived density and temperature ranges are listed in 
Table~\ref{table:DTlist}. In the case of $\eta$~Tau, we chose to list 
only an upper limit for the density, since the lower limit is 
essentially zero. 
 
\section{CH Production in Moderately Dense Regions \label{section:EQC}}

To estimate the amount of CH produced in the denser regions in 
our sight lines, we followed
the method described in detail by Federman et al. (1997). In their work, they 
only used the most
important processes to find the analytic expression
\begin{equation}
x({\rm CH})= \; \frac{0.67 \, k_1 \, x({\rm C}^+) \, x({\rm H}_2) \, n}
{G({\rm CH})} \label{eq:N(CH)}
\end{equation}
that describes the CH fractional abundance. In eqn.~\ref{eq:N(CH)} and 
all the following equations, 
$k_i$ denotes the rate coefficient for reaction i and the fractional 
abundance of a species y is $x$(y)= $n$(y)/($n$(H) + 2$n$(H$_2$)). 
The terms $n$, $n$(y), and $G$(y) are the 
gas density, number density
of species y, and the photodissociation rate for species y, respectively. The 
photodissociation rate 
includes grain attenuation of ultraviolet radiation which is 
approximated by $\tau_{UV}$ = 2$A_V$ 
(Federman et al. 1994). For the sight lines in Scorpius, which have larger 
ratios of total to selective extinction, a value of 1.4$A_V$ was adopted.  
One can invert eqn.~\ref{eq:N(CH)} and solve for $n$ using
the observed CH abundance 
for a comparison between the densities derived from \ion{C}{1} excitation.
We can also constrain our results for CH, using available upper limits for 
$N$(C$_2$) and $N$(CN). 
In equilibrium, the fractional abundances are given by the following 
expressions (Federman et al. 1997): 
\begin{equation}
x({\rm C}_2)= \; \left( \frac{1}{0.67} \right) \left( \frac{k_2}{k_1} \right) 
\left( \frac{G({\rm CH})}{G({\rm C}_2)} \right) 
\left( \frac{x({\rm CH})}{x({\rm H}_2)} \right) \, x({\rm CH}) 
\label{eq:N(C2)withN(CH)}
\end{equation}
and
\begin{equation}
x({\rm CN})= \left( \frac{1}{0.67} \right) 
\left( \frac{k_4 x({\rm N}) + 0.045 k_5 \, x({\rm C}^+)}
{k_1 x({\rm C}^+)} \right) 
\left( \frac{G({\rm CH})}{G({\rm CN})} \right) 
\left( \frac{x({\rm CH})}{x({\rm H}_2)} \right) x({\rm CH}) \; . 
\label{eq:N(CN)withN(CH)}
\end{equation}
The rate coefficients and photodissociation rates used in this analysis 
are given in Table~\ref{table:MCreactions}. 
A detailed description of these coefficients is 
given by Federman et al. (1994) with updates in Pan, Federman, \& Welty 
(2001). Table~\ref{table:MCdata} lists the necessary fractional 
abundances and $\tau_{UV}$ calculated from 
$E$($B-V$) values of Savage et al. (1977). The data for H$_2$ and the 
carbon-bearing molecules, CH, C$_2$, and CN, 
are from Savage et al. (1977) and Federman et al. (1994), respectively. 
The CO abundances are from recent analyses of $Copernicus$ spectra 
(Crenny \& Federman 2001, private communication).
We used the CH$^+$ abundances of Federman (1982), Lambert \& Danks (1986), 
and Price et al. (2001).  The data for $x$(C$^+$) are from Sofia et al. 
(1997), $x$(O) from Meyer, Jura, \& Cardelli (1998), and $x$(N) 
from Lugger et al. (1978) and York et al. (1983).

We estimated the gas densities necessary to reproduce the 
observed CH columns in an 
environment in thermal equilibrium by inverting eqn.~\ref{eq:N(CH)} and 
using the data in Tables~\ref{table:MCreactions} and \ref{table:MCdata}. 
Such environments are very different from
those required to produce the observed amount of CH$^+$. The calculated 
upper limits on densities, as expected, are much larger 
than those found from \ion{C}{1} excitation (see Section \ref{table:DTlist}). 
The sole exception is 
$\sigma$~Sco, where the two densities are marginally consistent with 
each other. Since we chose lines of 
sight with no detectable C$_2$ and CN absorption, densities of several thousand 
cm$^{-3}$ are not 
realistic, and the values inferred from \ion{C}{1} excitation are 
more appropriate. The 
clear inconsistency resulting from the two density calculations is a 
strong indication that 
non-equilibrium CH chemistry is important toward these stars. This 
contrasts with the results for 
the molecule-rich gas toward $\zeta$~Oph (Lambert et al. 1994) where the above
calculation and the observations gave similar results.

Using $x$(C$_2$) or $x$(CN) as well as $N_{total}$ from 
Table~\ref{table:MCdata}, we can
estimate the column densities of CH associated with the equilibrium 
chemistry from eqns.~\ref{eq:N(C2)withN(CH)} or \ref{eq:N(CN)withN(CH)}. 
Then, a lower limit on the amount 
of CH associated with the non-equilibrium CH$^+$ chemistry can be 
imposed by comparing the 
calculated amounts to the observed column densities. 
On the other hand, if we use the densities derived from \ion{C}{1} 
excitation in eqn.~\ref{eq:N(CH)}, 
an upper limit on the amount of CH produced via CH$^+$ can be obtained. In 
Figure~\ref{thermal_chem} 
we display the results, normalized to the observed column 
of CH$^+$. The triangles and 
downward-pointing arrows represent the measured values and upper limits of the 
$N$(CH)/$N$(CH$^+$) ratios, 
respectively. Ranges indicate the normalized CH columns that are likely to be 
associated with 
the non-equilibrium CH$^+$ chemistry. Again, the results in 
Figure~\ref{thermal_chem} strongly suggest 
that non-equilibrium chemistry is indeed very important for the 
production of CH in molecule-poor sight 
lines, and in some cases, like that of 23~Ori, it is the dominant 
source for this molecular species. 

A larger contribution from the non-equilibrium CH$^+$ chemistry would be 
necessary if CH is effectively destroyed by reactions with 
atomic hydrogen (reaction 30 in Table~\ref{table:reactions}) in the moderately 
dense regions along our sight lines. This reaction is thought to
be suppressed by a reaction barrier (e.g., Mitchell \& Deveau 1983); 
however, recent quantal calculations (Harding, Guadagnini, \& Schatz 1993; 
van Harrevelt, van Hemert, \& Schatz 2002) suggest that this may not be the case. 
Unfortunately, the rate coefficient for the reaction is not well defined 
for the gas temperatures relevant to the diffuse ISM, and the value of $x$(H) 
in the moderately dense regions is not known $-$ only line of sight averages 
are available. Therefore, the importance of the reaction, CH + H $\rightarrow$ 
C + H$_2$, could not be judged properly and the reaction was not included in eqns.
\ref{eq:N(CH)} $-$ \ref{eq:N(CN)withN(CH)}. If significant amounts of CH are 
destroyed through this reaction,
our conclusion about the importance of non-equilibrium chemistry would
be strengthened.

\section{Nonthermal Chemistry in Diffuse Clouds \label{section:NEQC}} 

We used the chemical model of Federman et al. (1996a) to examine the effects of 
nonthermal 
motions on diffuse cloud chemistry. The origin of the nonthermal 
motion is assumed to be the result of the 
propagation of Alfv\'{e}n waves, constantly entering the cloud from the 
intercloud medium. 
The dissipation of these waves is concentrated primarily in a diffuse 
cloud boundary layer, the 
transition zone between the cloud and the intercloud medium. The 
molecular abundances resulting 
from the Alfv\'{e}nic driven nonthermal chemistry are calculated for 
this boundary layer, without 
considering any time dependence or attempting to describe any 
changes in physical or chemical 
conditions throughout the region. Instead, we impose a set of 
conditions for the clouds toward 
our sample of stars that are inferred from previous observations or 
are from the excitation 
analysis of neutral carbon. 

In the chemical model we use a set of 16 atomic and molecular species closely 
related to CH$^+$ 
chemistry. These are H, He, C, O, He$^+$, C$^+$, H$_2$, CO, CH, OH, H$_2$O, 
CO$^+$, CH$^+$, 
CH$_2^+$, CH$_3^+$, and HCO$^+$. The set of molecules and chemical 
reactions was chosen so that it 
constitutes a closed system with reasonable assumptions on the 
physical conditions in diffuse clouds. 
The chemical reactions involved in this model are given in 
Table~\ref{table:reactions}; the rate 
coefficients were taken from Federman et al. (1997) and Draine \& Katz (1986), 
updated with information from van Dishoeck \& Black (1988), 
Federman \& Huntress (1989), Mitchell (1990), Anicich (1993), Millar, Farquhar, 
\& Willacy (1997), and Pan et al. (2001).  
For example, the CO photodissociation rates come from 
the tabulation in van Dishoeck \& Black (1988) using the observed column 
densities of CO and H$_2$ and $\tau_{UV}$.  The photodissociation 
rates for the other molecules included grain attenuation through
the $\tau_{UV}$ values listed in Table~\ref{table:MCdata}. 
Abundances of CH, CO, CO$^+$, HCO$^+$, He$^+$, CH$_2^+$, and CH$_3^+$ 
were calculated by solving a
system of nonlinear rate equations. We assumed that for all 
lines of sight the He, OH, 
and H$_2$O fractional abundances are 0.1, 1.6 $\times$ 10$^{-10}$, and 
3.2 $\times$ 10$^{-11}$, 
respectively, based on the theoretical results of Federman et al. (1996b). 
The abundances of the remaining elements and molecular species, listed in 
Table~\ref{table:MCdata}, 
were taken from observations (see Section~\ref{section:EQC}) and were 
kept fixed throughout the analysis.  Fractional abundances of neutral 
carbon are from the present analysis.  The 
electron densities were derived from the C$^+$ abundances.

We define an effective temperature for each ion-neutral (and in some 
cases for each neutral-neutral) reaction in order to introduce the 
effects of nonthermal motions. Federman et al. (1996a) defined $T_{eff}$ as

\begin{equation}
\frac{3}{2}k T_{eff}= \; \frac{3}{2} k T \; + \; \frac{1}{2} 
\mu v^2_{particle} \; .
\end{equation}
where $k$ and $\mu$ are the Boltzmann constant and the reduced 
mass of the system, respectively. The term $v_{particle}$ is the 
turbulent velocity of either 
the ionic ($v_{ion}$) or the neutral ($v_{neutral}$) species and 
defined by the root mean square 
Alfv\'{e}n wave amplitude or zero, depending on the physical environment we 
attempted to model.

First, we tested our code by attempting to reproduce the results of Federman 
et al. (1996a). 
They imposed a set of conditions on their model clouds that are believed to be 
representative of 
the region in which the effects of nonthermal chemistry are 
significant. Therefore, we modeled 
their ``hypothetical'' cloud with fractional abundances of 0.2, 
0.076, 2 $\times$ 10$^{-4}$, 
7.94 $\times$ 10$^{-5}$, and 3.02 $\times$ 10$^{-4}$ for H$_2$, He, C$^+$, N, 
and O, respectively. 
The visual extinction was set to 0.3 mag, representative of a 
predominantly atomic diffuse cloud, 
corresponding to a value of 0.6 for $\tau_{UV}$. Since we did not 
calculate CH$^+$ abundances, 
as opposed to Federman et al. (1996a), we used their result for CH$^+$ to 
calculate the 
amounts of CO$^+$, HCO$^+$, and CH produced by the nonthermal chemistry. The 
gas density, temperature, and $v_{ion}$ combinations used for the modeling 
correspond to their models 1, 2 and 3 $-$ (100 cm$^{-3}$, 100~K, 
3~km~s$^{-1}$), (50 cm$^{-3}$, 100~K, 3~km~s$^{-1}$), and
(200 cm$^{-3}$, 100~K, 3~km~s$^{-1}$), respectively.  
In all cases the calculated CO$^+$, HCO$^+$, and CH abundances were 
within 30--40\% of the values 
listed in Table~1 of Federman et al. (1996a). 

In the models of Federman et al. (1996a), the nonthermal effects were 
considered only for 
charged species, reflecting the distinction that ions are tied to the 
magnetic field. However, if 
one considers the conservation of momentum during a chemical reaction, 
then it is plausible that 
the product species (ion or neutral) could carry some of the momentum from the 
parent molecules (Gredel et al. 1993; Bucher \& Glinski 1999). 
This would mean that even neutral species, directly produced by 
reactions involving ionic species, 
can have more turbulent velocity than the average value for neutrals 
(not related directly to the 
ionic species). One could establish a complex model, where the 
turbulent velocity progressively 
decreases from the value representative for ions to that of the 
average for neutrals, following 
the chain of chemical reactions from ionic to neutral species. 
Such an approach is beyond the 
scope of this study. Instead, we simply explored the phenomena by 
setting the turbulent velocity 
for neutral molecules either to the ionic turbulent velocity or zero. 

In our approach, we modeled three different situations, one with 
no turbulence at all (e.g. 
equilibrium chemistry), one with $v_{ion} \neq$ 0 but $v_{neutral}$ $=$ 0, 
and one with nonzero 
$v_{ion}$ and $v_{neutral}$ $=$ $v_{ion}$. We used the equilibrium CH$^+$ 
abundance of 3.3$\times$10$^{-12}$ 
calculated by Federman et al. (1996a) whenever we attempted to model 
equilibrium chemistry without 
any turbulence. Since the observed CH$^+$ abundance is primarily the product
of non-equilibrium chemistry, we would introduce the effects of this 
chemistry into our ``equilibrium'' run 
($v_{ion}$ $=$ 0 and $v_{neutral}$ $=$ 0) by using the observed amount 
of CH$^+$. However, we used the observed
amount of CH$^+$ in the cases when the turbulent velocities were 
set to nonzero. The values for
turbulent velocity were chosen to be the amplitude of Alfv\'{e}nic waves. 
We adopted a standard 
value of 3~km~s$^{-1}$ for both $v_{ion}$ and $v_{neutral}$ (whenever 
they differ from zero),
following Federman et al. (1996a), based on the typical width of 
CH$^+$ absorption lines. We used the temperature and density from the
\ion{C}{1} excitation analysis (see Table~\ref{table:DTlist}) 
as $T$ and $n$.

The results of our calculations are shown in Tables~\ref{table:CHfromnEQ} and 
\ref{table:COfromnEQ}
for CH and  CO. It is immediately obvious that the observed abundances 
cannot be reproduced by 
equilibrium chemistry alone; there are several orders of magnitude 
difference between the observed 
and calculated abundances for all lines of sight for both CH and 
CO. Our results indicate that a 
nonthermal chemical model for ionic species only produces too 
much CH. When we also take nonthermal 
effects into account for neutral--neutral reactions, 
the observed CH abundances are reproduced reasonably well.  The 
lone exception is the unusual line of sight toward 23 Ori, which has 
detectable amounts of CH and CH$^+$ even though the H$_2$ column is 
about 10$^{18}$ cm$^{-2}$.  When $T_{eff}$ increases, reaction 17 
(see Table~\ref{table:reactions}) becomes more effective in destroying CH. This 
reaction also 
increases the amount of CO between models 2 and 3. 
Unlike the situation for CH, we did not find reasonable 
agreement for CO in general, indicating the limits of our 
approximations.   The closest correspondence occurs for the denser sight 
lines, those in Scorpius, suggesting that pockets of higher density control 
the abundance of CO in diffuse gas.  
Since a theoretical determination of the rate coefficient for 
C$^+$ $+$ OH $\rightarrow$ CO$^+$ $+$ H (Dubernet, Gargaud, \& 
McCarroll 1992) suggests a value much larger than that used here, 
we investigated its affect on CO as well.  While the equilibrium 
abundance for CO increases nearly 5 fold, the non-equilibrium 
results are changed only 5\%.  Clearly, the production of CO$^+$ 
and HCO$^+$, the precursors of CO, is dominated by the presence 
of enhanced amounts of CH$^+$ brought about by non-equilibrium 
chemistry in the sight lines considered here. It is likely that we 
have to improve our model in other ways as well.  For instance, a more 
precise description for $v_{neutral}$ is needed as is 
a more detailed analysis of H$_2$ and CO photodissociation
that includes self- and mutual--shielding.

\section{Discussion \label{section:Discussion}}

Observations and models have suggested long ago that the ISM is not 
in an equilibrium state. The underlying reason is that interstellar 
clouds cannot be treated as a closed systems; interactions with often 
violent surroundings are inevitable. Shock fronts from supernovae and 
cloud collisions are probably commonplace. These non-equilibrium 
processes affect the chemistry of most molecular species, including 
that of CH$^+$. In fact, almost everyone agrees that these 
non-equilibrium processes hold the key for explaining the elusive 
CH$^+$ abundance problem in the ISM. The exact mechanisms and place
for CH$^+$ production, however, is still the subject of intense 
debate. MHD shock models (e.g., Flower \& Pineau des For\^{e}ts 1998) or
those that involve intermittent dissipation of interstellar turbulence
(e.g., Joulain et al. 1998) suggest that CH$^+$ is formed in warm 
material at temperatures $T >$100 K. Excitation analysis of 
interstellar C$_2$ (Gredel 1999), on the other hand, places the 
production of CH$^+$ into the colder gas at temperatures of 50-100 K. 
Our results are more
consistent with this later suggestion and support the claim
that non-equilibrium chemistry is very important (or 
sometimes even dominant) for CH$^+$, CH, and CO in diffuse sight lines.
We concur with Gredell (1999) that about 30\% of the observed CH 
is associated with CH$^+$ synthesis.  
The associated molecular ions, CO$^+$ and HCO$^+$, are likely to 
be affected as well, with respective abundances reaching 
$\sim 6 \times 10^{-13}$ and about $5 \times 10^{-11}$.  (We note 
in passing that such enhancements in the HCO$^+$ abundance still 
yield column densities at least an order of magnitude below those 
inferred from absorption of the millimeter-wave continuum from 
extragalactic sources [Lucas \& Liszt 1996].). 
This conclusion highlights the fact that any model attempting to 
account for CH$^+$ abundances must address the ``side-effects'' on 
other species and be able to 
describe their abundances. Therefore, indirectly, we imposed constraints on the 
suggested theories thought to be responsible for CH$^+$ production in 
this environment.

There were cases where our simplified models failed, indicating some 
shortcomings in our approach. It 
remains to be seen whether the source of these discrepancies is related to the 
assumed physical 
environment for particular lines of sight (gradients in physical conditions) 
or to effects ignored in our approach, such as detailed 
calculations for H$_2$ and CO photodissociation. Another possible 
source of error is the 
``uniformity'' of our approximations. The differences in $b$-values 
and Doppler shifts associated with the \ion{C}{1} fine structure levels 
for a given sight line indicate that the levels 
might not fill the same volume. The excitation may occur in clumps.  
A more careful excitation analysis of neutral carbon is 
necessary to infer average density and temperature for diffuse clouds, 
but the means to perform 
such an analysis are beyond the capabilities of 
current theory and observation.  Moreover, we either 
applied a given turbulence for neutrals or we did not. In reality, the 
turbulence can change 
gradually from the values representative for ions to that of neutrals as 
one follows the chemical 
network from an ionic parent molecule (like CH$^+$ in our case) to 
products (Bucher \& Glinski 1999). Further work 
should explore the importance of these various effects. 

\section{Summary \label{section:Summary}}

We surveyed the physical and chemical conditions in the diffuse ISM with the 
aim of assessing the importance of non-equilibrium processes.
We retrieved a mixture of high and medium resolution spectra 
from the $HST$ archive. The selection criteria for the lines of sight 
was motivated by the idea to minimize the contribution from the denser 
diffuse cloud cores, which are satisfactorily described by equilibrium 
chemistry. We studied  transitions involving the fine structure 
levels of neutral carbon to infer the physical parameters of these 
clouds through analysis of excitation. Our conclusions can be 
summarized as follows:

We could identify multiple cloud components along the sight lines
for which high-resolution spectra were available and toward HD~112244
where the separation between the components was larger than the resolution
of the MR spectra. We derived \ion{C}{1} column densities and $b$-values 
for the components or provided upper limits for non-detections.

We performed \ion{C}{1} excitation analyses for all main components 
to infer $n$ and $T$. The gas densities and kinetic 
temperatures were comparable to the representative values for 
diffuse clouds.  

We calculated the abundances produced by molecular chemistry in the 
moderately dense
regions along our sight lines. By removing the contribution from the 
molecular chemistry, we quantified the amounts of CH from nonthermal 
chemistry. Our analysis suggested that the amount of CH associated 
with CH$^+$ is at least 30--40\% of the amount of CH in general, 
but in some cases it can reach as high as 90\% (e.g., 23~Ori). 
 
Density estimates based on the observed upper limits for C$_2$ and CN, 
along with CH columns, and on the assumption that the equilibrium 
chemistry acts alone were (much) larger than the densities from 
\ion{C}{1} excitation. The discrepancy between these density estimates
clearly indicates that the assumption of equilibrium chemistry as the
sole contributor is inappropriate.

We modeled the non-equilibrium chemistry related to CH$^+$ and improved 
upon the approximation of Federman et al. (1996a) to account for nonthermal 
motions, the driving force behind non-equilibrium chemistry. We defined 
effective temperatures for both ions and neutrals which involved thermal 
and nonthermal motions. We explored the effect of these motions with 
different combinations of neutral and ion turbulent velocities. 
We found that the observed amounts of molecular species cannot 
be reproduced by equilibrium chemistry. Models with non-zero 
turbulence for ions only also fail; turbulence 
for both ionic and neutral species was required to match the 
observations satisfactorily. 

In closing, non-equilibrium chemistry is needed 
to explain the physical and chemical conditions in the diffuse gas 
toward our sample of stars and substantial amounts of CH are produced by 
nonthermal chemistry in the diffuse ISM along these sight lines.  

\begin{center}
{\bf A. Appendix}
\end{center}

Here we present tables of equivalent widths for C~{\small I} lines seen 
in our spectra (Tables~\ref{table:App1}-\ref{table:App2}).  
Comparison is made with other determinations, when 
available, and with the results of our profile syntheses.  The agreement 
among the various measures is quite good, especially when one considers the 
fact that the MR spectra from $HST$ and the $Copernicus$ spectra do not 
always resolve blended lines or velocity components.  Specific examples 
are given in the notes to each table.

\acknowledgements

We thank Dan Welty for his comments on the Ph.D. Dissertation from the 
Univ. of Toledo on which this paper is based. Comments by the anonymous
referee improved several points in our paper. 
This research was supported by STScI grant AR-08352.01-97A.  For the 
$Copernicus$ data, we utilized the archive developed by George Sonneborn 
and available at the Multiwavelength Archive at STScI.  We made use of the 
Simbad database, operated at Centre de Don\'{e}es Astronomiques de 
Strasbourg, Strasbourg, France.
 
\clearpage
\begin{deluxetable}{lccccccc}
\tablecolumns{8}
\tabletypesize{\footnotesize}
\tablewidth{0pc} 
\tablecaption{Data for Target Stars \label{table:LOS}}
\tablehead{
HD     &     Star        & \multicolumn{2}{c}{Galactic 
Coordinates} & Distance & Spectral type & $V$ & $E$($B-V$) \\ \cline{3-4}
       &                 &    $l$ ($^{\circ}$)   &    $b$ 
       ($^{\circ}$)  &    (pc)      &   & (mag) & (mag) \\
}
\startdata
23630  &   $\eta$~Tau    &  166.67  &  -23.46  &   
113    &       B7IIIe    & 2.87 & 0.00 \\
24760  &  $\epsilon$~Per &  157.35  &  -10.09  &   
165    &       B0.5V     & 2.90 & 0.09 \\
36861  &  $\lambda$~Ori  &  195.05  &  -12.00  & 
$\ldots$  &      O8III   & 3.30 & 0.12 \\
112244 &   $\ldots$      &  303.55  &    6.03  &   
578    &       O9Ibe     & 5.32 & 0.34 \\
141637 &      1~Sco      &  346.10  &   21.71  &   
160    &       B1.5Vn    & 4.64 & 0.20 \\
143018 &  $\pi$~Sco      &  347.21  &   20.23  &   
141    &       B1V       & 2.89 & 0.08 \\
143275 &   $\delta$~Sco  &  350.10  &   22.49  &   
123    &       B0.2IVe   & 2.29 & 0.16 \\
144470 &  $\omega^1$~Sco &  352.75  &   22.77  &   
130    &       B1V       & 3.95 & 0.22 \\
147165 &   $\sigma$~Sco  &  351.31  &   17.00  &   
225    &       B1III     & 2.91 & 0.38 \\
\enddata
\end{deluxetable}

\clearpage
\begin{deluxetable}{ll}
\tablecolumns{2}
\tabletypesize{\footnotesize}
\tablewidth{0pc} 
\tablecaption{$HST$ Observations for Our Sight Lines 
\label{table:Obs}}
\tablehead{  
\colhead{Star} & 
\colhead{Dataset} \\
}
\startdata
$\eta$~Tau & z1bw020st,z1bw020tt,z1bw020ut,z1bw020vt,z1bw020wt, \\
 & z1bw020yt,z1bw0210t \\
$\epsilon$~Per & z2c20206t,z2c20207t,z2c20208t \\ 
$\lambda$~Ori  & z3di0204t,z3di0205t,z2c20306t,z2c20307t,z2c20308t, \\
 & z2520306t,z2520307t,z2520308t \\
HD~112244      & z0yv0506m,z0yv050gt,z0yv060em \\
1~Sco          & z2zx020bp,z2zx020fp,z0zi0708m,z0zi070bm,z0zi070ct, \\
 & z0zi070dt,z0zi070gt,z0zi070it \\
$\pi$~Sco      & z0zi0408t,z0zi0409t,z0zi040at,z0zi040bt,z0zi040ct, \\
 & z0xz0308t,z0xz0309t,z0xz030at,z0zi040et,z0zi040gt \\
$\delta$~Sco  & z3di0504t,z3di0505t,z34r020dt,z34r020et,z2c20506t, \\
 & z2c20507t,z2c20508t \\
$\omega^1$~Sco & z2c20606t,z2c20607t,z2c20608t \\
$\sigma$~Sco   & z34r0409t,z13v0108m,z13v0109m,z13v010bm,z13v010cm, \\
 & z13v010dm,z13v010gm \\
\enddata      
\end{deluxetable}

\clearpage
\begin{deluxetable}{lcccccc}
\tablecolumns{7}
\tabletypesize{\footnotesize}
\tablewidth{0pc} 
\tablecaption{Derived Parameter Values \label{table:fit}}
\tablehead{Star & \multicolumn{2}{c}{Component A} & 
 & \multicolumn{3}{c}{Component B} \\ \cline{2-3} \cline{5-7}
Level & Column density & $b$-value & & Doppler shift$^a$ & 
Column density & $b$-value \\
 & (cm$^{-2}$) & (km s$^{-1}$) & & (km s$^{-1}$) & 
 (cm$^{-2}$) & (km s$^{-1}$) \\
}
\startdata
$\eta$~Tau & & & & & & \\       
 0 & 2.4($\pm$0.2) $\times$ 10$^{13}$ & 2.5$\pm$0.3 
 & & $\ldots$ & $\ldots$ & $\ldots$ \\
 1 & $\leq$ 7.7 $\times$ 10$^{11}$ & $\ldots$ 
 & & $\ldots$ & $\ldots$ & $\ldots$ \\
 2 & $\leq$ 5.2 $\times$ 10$^{11}$ & $\ldots$ 
 & & $\ldots$ & $\ldots$ & $\ldots$ \\
 & & & & & & \\
$\epsilon$~Per & & & & & \\
 0 & 4.4($\pm$0.4) $\times$ 10$^{13}$ & 3.0$^b$ 
 & & $\ldots$ & $\ldots$ & $\ldots$ \\
 1 & 6.4($\pm$0.6) $\times$ 10$^{12}$ & 
2.1$\pm$0.2 & & $\ldots$ & $\ldots$ & $\ldots$ \\
\enddata
\tablenotetext{a}{ The quantity ``Doppler shift'' is a shift relative to 
Component A. It was measured in HR (FWHM~$\approx$~3 km s$^{-1}$) 
spectra and was sometimes adjusted during profile synthesis to improve the fit.
In the case of HD~112244, the large shift between the two component
could be measured in MR (FWHM~$\approx$~15 km s$^{-1}$) spectra.}
\tablenotetext{b}{ The parameter was kept fixed during the fitting procedure.}
\end{deluxetable}

\clearpage
\begin{deluxetable}{ccccccccc}
\tablecolumns{9}
\tabletypesize{\footnotesize}
\tablewidth{0pc} 
\tablecaption{Collisional deexcitation coefficients 
\label{table:collision}}
\tablehead{Coeff.  & & \multicolumn{5}{c}{Transition 
from level i to level j} & & Notes \\ \cline{3-7}
 (cm$^3$ s$^{-1}$) & &  $J$=2 $\rightarrow$ 0  & &  
$J$=2 $\rightarrow$ 1  & &  $J$=1 $\rightarrow$ 0 & & \\
}
\startdata
                       & &                       & &                       
                       & &                       & & \\
$\gamma^{H}_{i,j} $    & & 7.244 $\cdot$ 10$^{-11}$ 
T$^{0.069}$ & & 1.413 $\cdot$ 10$^{-10}$ T$^{0.156}$ 
& & 1.349 $\cdot$ 10$^{-10}$ T$^{0.043}$ & & 1 \\
                       & &                       & &                       
                       & &                       & & \\
$\gamma^{He}_{i,j}$    & & 1.995 $\cdot$ 10$^{-11}$ 
T$^{0.181}$ & & 2.951 $\cdot$ 10$^{-11}$ T$^{0.241}$ 
& & $\approx$ 0           & & 2,3 \\
                       & &                       & &                       
                       & &                       & & \\
$\gamma^{ortho-H_2}_{i,j}$  & & 3.311 $\cdot$ 10$^{-11}$ 
T$^{0.16 }$ & & 4.365 $\cdot$ 10$^{-11}$ T$^{0.27 }$ 
& & 8.511 $\cdot$ 10$^{-11}$ T$^{-0.03}$ & & 4 \\
                       & &                       & &                       
                       & &                       & & \\
$\gamma^{para-H_2}_{i,j}$  & & 1.072 $\cdot$ 10$^{-10}$ 
T$^{-0.04}$ & & 1.479 $\cdot$ 10$^{-10}$ T$^{0.05 }$ 
& & 1.122 $\cdot$ 10$^{-10}$ T$^{-0.10}$ & & 4 \\
                       & &                       & &
                       & &                       & & \\
\enddata
\tablenotetext{~}{ The quantity $T$ is the gas temperature.} 
\tablenotetext{~}{Notes. $-$ (1) Launay \& Roueff 1977; 
(2) Lavendy et al. 1991; (3) Staemmler \& Flower 1991; (4) 
Schr\"{o}der et al. 1991.}
\end{deluxetable}

\clearpage
\begin{deluxetable}{lccccccc}
\tablecolumns{8}
\tabletypesize{\footnotesize}
\tablewidth{0pc} 
\tablecaption{Gas Densities and Temperatures  
\label{table:DTlist}}
\tablehead{
Star                         & &    
$n_{C~I}$$^a$ 
& &   $T_{H_2}^b$    & & & $n$$^c$ \\
                             & &        (cm$^{-3}$) 
                             & &   (K)    & & &     
                             (cm$^{-3}$)     \\
}
\startdata
$\eta$~Tau                   & &          $\leq$3.5  
& &  59--143 & & &       $\ldots$       \\
$\epsilon$~Per               & &           11.5--33 
& &  51--188 & & &       $\ldots$       \\
$\lambda$~Ori                & &            68--182 
& &  36--60  & & &       $\ldots$       \\
HD~112244~A$^d$ & &             14--28 
& &  67--113 & & &      $\leq$  960     \\
HD~112244~B$^d$ & &           16.5--48 
& &  67--113 & & &      $\leq$  960     \\
1~Sco                        & &           200--800  
& &  53--114 & & &      $\leq$ 2299     \\
$\pi$~Sco                    & &             13--33  
& &  56--144 & & &      $\leq$ 2793     \\
$\delta$~Sco                 & &            90--265 
& &  42--81  & & &      $\leq$ 1115     \\
$\omega^1$~Sco               & &             43--83 
& &  59--94  & & &      $\leq$ 643      \\
$\sigma$~Sco                 & &           137--345 
& &  50--87  & & &      $\leq$ 371      \\
23~Ori                       & &            7.5--23  
& &  65--150 & & &      $\leq$ 4004     \\ 
$\beta^1$~Sco                & &            35--205  
& &  77--103 & & &      $\leq$ 770      \\
\enddata
\tablenotetext{a}{ Density inferred from the \ion{C}{1} level populations.}
\tablenotetext{b}{ H$_2$ excitation temperatures of Savage et al. 1977.}
\tablenotetext{c}{ Density calculated by inverting eqn.~\ref{eq:N(CH)}.}
\tablenotetext{d}{ A is the stronger and B is weaker component.}
\end{deluxetable}

\clearpage
\begin{deluxetable}{lclclcc}
\tablecolumns{5}
\tabletypesize{\footnotesize}
\tablewidth{0pc} 
\tablecaption{Chemical Network for the Equilibrium Chemistry 
\label{table:MCreactions}}
\tablehead{
Reaction & & Rate coefficient & & Value & & Notes \\
  & & or Rate  &  & & & \\
}
\startdata
C$^+$ + H$_2$ $\rightarrow$ CH$^+_2$ + h$\nu$ & & $k_1$ 
& & 1.0 $\times$ 10$^{-16}$ cm$^3$ s$^{-1}$ & & 1 \\
C$^+$ + CH $\rightarrow$ C$^+_2$ + H          & & $k_2$ 
& & 5.4 $\times$ 10$^{-10}$ cm$^3$ s$^{-1}$ & & 2 \\
C$_2$ + N $\rightarrow$ CN + C                & & $k_3$ 
& & 1.7 $\times$ 10$^{-11}$ $\sqrt{\frac{T}{300}}$ cm$^3$ s$^{-1}$ & & 3 \\
CH + N $\rightarrow$ CN + H                   & & $k_4$ 
& & 2.0 $\times$ 10$^{-11}$ $\sqrt{\frac{T}{300}}$ cm$^3$ s$^{-1}$ & & 4 \\
C$^+$ + NH $\rightarrow$ $\left\{ \begin{array}{c}
                                  \mbox{CN$^+$ + H} \\
                                  \mbox{CH$^+$ + N} 
                                  \end{array} \right.$   
              & & $k_5$ & & 5.0 $\times$ 10$^{-10}$ cm$^3$ s$^{-1}$ & & 3 \\
CH + h$\nu$ $\rightarrow$ C + H               & & 
$G$(CH) & & 1.3 $\times$ 10$^{-9}$ e$^{-\tau_{UV}}$ s$^{-1}$ & & 2 \\
C$_2$ + h$\nu$ $\rightarrow$ 2C               & & 
$G$(C$_2$) & & 2.0 $\times$ 10$^{-10}$ e$^{-\tau_{UV}}$ s$^{-1}$ & & 5 \\
CN + h$\nu$ $\rightarrow$ C + N               & & 
$G$(CN) & & 1.0 $\times$ 10$^{-10}$ e$^{-\tau_{UV}}$ s$^{-1}$ & & 6 \\
\enddata
\tablenotetext{~}{Notes.-- (1) Federman et al. 1997; (2) Pan et al. 2001; 
(3) Federman et al. 1994; (4) Messing et al. 1981; (5) Pouilly et al. 1983; 
(6) Lavendy et al. 1987.}
\end{deluxetable}

\clearpage
\begin{deluxetable}{lcccccccccccc}
\rotate
\tablecolumns{13}
\tabletypesize{\footnotesize}
\tablewidth{0pc} 
\tablecaption{Fractional Abundances \label{table:MCdata}}
\tablehead{
 & & & \multicolumn{10}{c}{$x$(Y)$^a$} \\ \cline{4-13}  
\colhead{Star} & 
\colhead{$\tau_{UV}$} & 
\colhead{$N_{total}$$^b$} & 
\colhead{C$^+$} & 
\colhead{C} & 
\colhead{N} & 
\colhead{O} & 
\colhead{H$_2$} & 
\colhead{CH$^+$} & 
\colhead{CO} &
\colhead{CH} &
\colhead{C$_2$} &
\colhead{CN} \\
}
\startdata
$\eta$~Tau     & 0.00 & 2.7(20)$^c$ & 1.4(-4) & 9.1(-8) & 
7.5(-5) & 4.0(-4) & 2.6(-1) & 7.0(-9) & $\leq$8.1(-9) & $\ldots$ & 
$\ldots$ & $\ldots$ \\
$\epsilon$~Per & 0.56 & 3.2(20) & 1.4(-4) & 1.5(-7) & 3.2(-5) & 3.0(-4) 
& 2.1(-1) 
& $\leq$3.4(-9) & 2.7(-9) & 6.2(-9) & $\ldots$ & $\ldots$ \\
$\lambda$~Ori  & 0.74 & 6.3(20) & 1.3(-4) & 3.2(-7) & 7.9(-5) & 3.2(-4) 
& 4.0(-2) 
& $\leq$1.1(-9) & 6.2(-9) & $\ldots$ & $\ldots$ & $\ldots$ \\
HD~112244~A    & 1.05 & 7.5(20) & 1.4(-4) & 3.8(-7) & 7.5(-5) & 4.0(-4) 
& 1.9(-1) 
& 1.0(-8) & 1.8(-8) & 4.0(-9) & $\ldots$ & $\leq$4.2(-10) \\
HD~112244~B    & 1.05 & 7.5(20) & 1.4(-4) & 3.0(-7) & 7.5(-5) & 4.0(-4) 
& 1.9(-1) 
& $\leq$3.0(-9) & 5.1(-9) & 2.8(-9) & $\ldots$ & $\leq$4.2(-10) \\
1~Sco          & 0.87 & 1.6(21) & 1.6(-4) & 6.4(-8) & 1.0(-4) & 3.5(-4) 
& 2.0(-2) 
& $\leq$8.8(-10) & $\leq$1.6(-9) & $\leq$1.1(-9) & $\ldots$ & 
$\leq$2.4(-10) \\
$\pi$~Sco      & 0.35 & 5.6(20) & 1.4(-4) & 1.9(-8) & 7.5(-5) & 3.5(-4) 
& 7.0(-2) 
& $\leq$2.0(-9) & 1.5(-9) & $\leq$8.9(-10) & $\leq$3.0(-9) & $\ldots$ \\
$\delta$~Sco   & 0.69 & 1.4(21) & 1.8(-4) & 1.2(-7) & 9.8(-5) & 4.2(-4) 
& 4.0(-2) 
& 2.4(-9) & 2.2(-9) & 1.6(-9) & $\ldots$ & $\leq$9.3(-11) \\
\enddata
\tablenotetext{a}{ Fractional abundance of species Y.}
\tablenotetext{b}{ $N_{total}$= $N$(H~{\small I}) + 2$N$(H$_2$).}
\tablenotetext{c}{ 2.7(20)= 2.7$\times$10$^{20}$.}
\end{deluxetable}

\clearpage
\begin{deluxetable}{llccrcc}
\tablecolumns{7}
\tabletypesize{\footnotesize}
\tablewidth{0pc} 
\tablecaption{Chemical Network \label{table:reactions}}
\tablehead{
 & Reaction & & & Rate coefficient or Rate & & Notes \\ 
 }
\startdata
1. & CO$^+$ + H  $\rightarrow$  CO + H$^+$		& & & 
7.5 $\times$ 10$^{-10}$ cm$^3$ s$^{-1}$ & & 1 \\ 
2. & HCO$^+$ + e$^-$  $\rightarrow$  CO + H		& & & 
2.4 $\times$ 10$^{-7}$ $\times$ $\left( \frac{300}{T_{eff}} 
\right) ^{0.69}$ cm$^3$ s$^{-1}$ & & 2 \\
3. & CO + He$^+$  $\rightarrow$  C$^+$ + O + He	& & & 
1.6 $\times$ 10$^{-9}$ cm$^3$ s$^{-1}$ & & 1 \\
4. & CO + h $\nu$  $\rightarrow$  C + O		& & &  
$G(CO)$\tablenotemark{a}~~s$^{-1}$ & & \nodata \\
5. & CH$^+$ + O  $\rightarrow$  CO$^+$ + H		
& & & 3.5 $\times$ 10$^{-10}$ cm$^3$ s$^{-1}$ & & 3 \\
6. & CO$^+$ + H$_2$  $\rightarrow$  HCO$^+$ + H	& & & 1.4 
$\times$ 10$^{-9}$ cm$^3$ s$^{-1}$ & & 3 \\
7. & CO$^+$ + e$^-$  $\rightarrow$  C + O		& & & 
1.0 $\times$ 10$^{-7}$ $\times$ $\left( \frac{300}{T_{eff}} 
\right) ^{0.46}$ cm$^3$ s$^{-1}$ & & 2 \\
8. & H$_2$ + He$^+$  $\rightarrow$  He + H$_2^+$	& & & 
1.0 $\times$ 10$^{-13}$ cm$^3$ s$^{-1}$ & & 3 \\
\enddata
\tablenotetext{~}{ Notes. $-$ (1) Draine \& Katz 1986; (2) Mitchell 1990; 
(3) Anicich 1993; (4) Millar et al. 1997; (5) Federman \& Huntress 1989;
(6) Pan et al. 2001; (7) Barsuhn \& Nesbit 1978; (8) Federman et al. 1997. }
\tablenotetext{a}{ $G$(CO) in units of 10$^{-10}$ s$^{-1}$ are 
1.6, 1.3, 1.2, 1.1, 1.1, 1.2, 1.4, 1.3, 1.1, 0.89, 1.6, and 1.1 
for $\eta$~Tau, $\epsilon$~Per, 
$\lambda$~Ori, HD~112244~A, HD~112244~B, 1~Sco, $\pi$~Sco, $\delta$~Sco, 
$\omega^1$~Sco, $\sigma$~Sco, 
23~Ori, and $\beta^1$~Sco, respectively.}
\end{deluxetable}

\clearpage
\begin{deluxetable}{lccccc}
\tablecolumns{6}
\tabletypesize{\footnotesize}
\tablewidth{0pc} 
\tablecaption{CH Fractional Abundances \label{table:CHfromnEQ}}
\tablehead{
Star & & Observed$^a$ & Model 1$^b$ & Model 
2$^c$ 
& Model 3$^d$ \\ 
 }
\startdata
$\eta$~Tau     & &           $\cdots$          & $\leq$6.1$\times$10$^{-12}$ 
& $\leq$2.8$\times$10$^{-9}$  & $\leq$2.7$\times$10$^{-9}$  \\
$\epsilon$~Per & &  $\leq$6.2$\times$10$^{-9}$ & (2.2-5.3)$\times$10$^{-11}$ 
& (0.5-1.2)$\times$10$^{-8}$  & (4.2-7.4)$\times$10$^{-9}$  \\
$\lambda$~Ori  & &            $\cdots$         & (1.2-1.8)$\times$10$^{-11}$ 
& (0.9-1.4)$\times$10$^{-9}$  & (4.6-5.2)$\times$10$^{-10}$ \\
HD~112244~A    & &     4.0$\times$10$^{-9}$    & (3.3-5.4)$\times$10$^{-11}$ 
& (2.2-3.6)$\times$10$^{-8}$  & (1.6-2.2)$\times$10$^{-8}$  \\ 
HD~112244~B    & &     2.8$\times$10$^{-9}$    & (3.6-6.9)$\times$10$^{-11}$ 
& (0.7-1.4)$\times$10$^{-8}$  & (5.2-7.4)$\times$10$^{-9}$  \\
1~Sco          & & $\leq$1.1$\times$10$^{-9}$  & (6.5-9.9)$\times$10$^{-12}$ 
& (3.5-5.3)$\times$10$^{-10}$ & (1.8-2.0)$\times$10$^{-10}$ \\
$\pi$~Sco      & & $\leq$8.9$\times$10$^{-10}$ & (0.6-1.3)$\times$10$^{-11}$ 
& (0.8-1.8)$\times$10$^{-9}$  & (0.7-1.1)$\times$10$^{-9}$  \\
$\delta$~Sco   & &     1.6$\times$10$^{-9}$    & (1.4-2.3)$\times$10$^{-11}$ 
& (1.9-3.0)$\times$10$^{-9}$  & (1.0-1.1)$\times$10$^{-9}$  \\
\enddata
\tablenotetext{a}{ References -- Federman 1982; Danks et al. 1984; Federman 
et al. 1997.}
\tablenotetext{b}{ Equilibrium chemistry ($v_{ion}$ = 0, $v_{neutral}$ = 0, 
$x$(CH$^+$) = 3.3$\times$10$^{-12}$).}
\tablenotetext{c}{ Non-equilibrium chemistry ($v_{ion}$ = 3 km s$^{-1}$, 
$v_{neutral}$ = 0,  observed $x$(CH$^+$)).}
\tablenotetext{d}{ Non-equilibrium chemistry ($v_{ion}$ = 3 km s$^{-1}$, 
$v_{neutral}$ = 3 km s$^{-1}$, observed $x$(CH$^+$)).}
\end{deluxetable}

\clearpage
\begin{deluxetable}{lccccc}
\tablecolumns{6}
\tabletypesize{\footnotesize}
\tablewidth{0pc} 
\tablecaption{CO Fractional Abundances \label{table:COfromnEQ}}
\tablehead{
Star & & Observed$^a$ & Model 1$^b$ & Model 
2$^c$ & 
Model 3$^d$ \\
}
\startdata
$\eta$~Tau     & & $\leq$8.1$\times$10$^{-9}$ & $\leq$7.7$\times$10$^{-13}$ 
& $\leq$1.6$\times$10$^{-10}$ & $\leq$1.7$\times$10$^{-10}$ \\
$\epsilon$~Per & & 2.7$\times$10$^{-9}$       & (2.8-6.9)$\times$10$^{-12}$ 
& (2.4-4.6)$\times$10$^{-10}$ & (2.6-5.8)$\times$10$^{-10}$ \\
$\lambda$~Ori  & & 6.2$\times$10$^{-9}$       & (1.2-3.2)$\times$10$^{-11}$ 
& (1.6-4.0)$\times$10$^{-10}$ & (1.8-4.4)$\times$10$^{-10}$ \\
HD~112244~A    & & 1.8$\times$10$^{-8}$       & (4.0-7.6)$\times$10$^{-12}$ 
& (1.1-1.8)$\times$10$^{-9}$  & (1.2-2.3)$\times$10$^{-9}$  \\ 
HD~112244~B    & & 5.1$\times$10$^{-9}$       & (0.5-1.3)$\times$10$^{-11}$ 
& (3.8-9.8)$\times$10$^{-10}$ & (0.4-1.2)$\times$10$^{-9}$  \\
1~Sco          & & $\leq$1.6$\times$10$^{-9}$ & (0.4-1.7)$\times$10$^{-10}$ 
& (0.3-1.2)$\times$10$^{-9}$  & (0.4-1.3)$\times$10$^{-9}$  \\
$\pi$~Sco      & & 1.5$\times$10$^{-9}$       & (2.4-5.7)$\times$10$^{-12}$ 
& (0.8-1.6)$\times$10$^{-10}$ & (0.8-1.8)$\times$10$^{-10}$ \\
$\delta$~Sco   & & 2.2$\times$10$^{-9}$       & (2.1-5.9)$\times$10$^{-11}$ 
& (0.5-1.3)$\times$10$^{-9}$  & (0.6-1.5)$\times$10$^{-9}$  \\
\enddata
\tablenotetext{a}{ Crenny \& Federman, private communication.}
\tablenotetext{b}{ Equilibrium chemistry ($v_{ion}$ = 0, $v_{neutral}$ = 0, 
$x$(CH$^+$) = 3.3$\times$10$^{-12}$).}
\tablenotetext{c}{ Non-equilibrium chemistry ($v_{ion}$ = 3 km s$^{-1}$, 
$v_{neutral}$ = 0, observed $x$(CH$^+$)).}
\tablenotetext{d}{ Non-equilibrium chemistry ($v_{ion}$ = 3 km s$^{-1}$, 
$v_{neutral}$ = 3 km s$^{-1}$, observed $x$(CH$^+$)).}
\end{deluxetable}

\clearpage
\clearpage
\begin{deluxetable}{ccc}
\tablecolumns{3}
\tablewidth{0pc} 
\tablecaption{$W_{\lambda}$ (m\AA) of \ion{C}{1} Lines toward 
$\eta$~Tau \label{table:App1}}
\tablehead{
Wavelength & Measured & Calculated \\
 (\AA)     &          &            \\ 
}
\startdata
J $=$ 0: &                  &       \\
1260.736 &  8.39 $\pm$ 1.51 & 10.64 \\
1277.245 & 16.38 $\pm$ 1.24 & 19.88 \\
1280.135 &  7.30 $\pm$ 1.09 &  7.38 \\
1328.833 & 16.59 $\pm$ 0.99 & 15.46 \\
1560.309 & 24.72 $\pm$ 0.86 & 24.50 \\
1656.928 & 33.87 $\pm$ 1.06 & 34.47 \\
\enddata
\end{deluxetable}

\begin{deluxetable}{ccccccc}
\tablecolumns{7}
\tablewidth{0pc} 
\tablecaption{$W_{\lambda}$ (m\AA) of \ion{C}{1} Lines toward 
$\epsilon$~Per}
\tablehead{
$\lambda$ & \multicolumn{2}{c}{Measured} & & 
\multicolumn{2}{c}{Previous Measurements} & Calculated 
\\ \cline{2-3} \cline{5-6}
 (\AA)    &    $HST$         &  $Copernicus$       & & 
 JS$^a$             &  JJL$^b$       &  \\ 
}
\startdata
J $=$ 0: &                  &                     & 
&                    &                    &          
\\
1260.736 & 18.77 $\pm$ 0.34 &  $\ldots$           & 
& $\ldots$           & 14.0 $\pm$ 0.7     &  17.32   \\
1328.833 & $\ldots$         &  22.3 $\pm$ 0.6     & & 
$\ldots$           & 29.0 $\pm$ 1.5     &  24.06   \\
J $=$ 1: &                  &                     & & 
                   &                    &          \\
1260.927 & $\leq$ 0.32      &  $\ldots$           & & 
$\ldots$           & $\leq$ 1.3         &   1.14   \\
1260.996 & $\leq$ 0.32      &  $\ldots$           & & 
$\ldots$           & $\leq$ 1.2         &   0.86   \\
1261.122 & $\leq$ 0.32      &  $\ldots$           & & 
$\ldots$           & $\leq$ 1.2         &   1.42   \\
1329.085 & $\ldots$         &  $\ldots$           & & 
$\ldots$           & $\ldots$           &   1.84   \\
\enddata
\tablenotetext{a}{ Jenkins \& Shaya 1979.}
\tablenotetext{b}{ Jenkins et al. 1983.}
\tablenotetext{c}{ Total $W_{\lambda}$ for $\lambda\lambda$1329.085, 
1329.100, 1329.123.}
\tablenotetext{d}{ Total $W_{\lambda}$ for $\lambda\lambda$1329.578, 
1329.601.}
\end{deluxetable}

\begin{deluxetable}{ccccccc}
\tablecolumns{7}
\tabletypesize{\footnotesize}
\tablewidth{0pc} 
\tablecaption{$W_{\lambda}$ (m\AA) of \ion{C}{1} Lines toward 
$\lambda$~Ori}
\tablehead{
$\lambda$ & \multicolumn{2}{c}{Measured} & &  
\multicolumn{2}{c}{Other Measurements}  & Calculated \\ 
\cline{2-3} \cline{5-6}
 (\AA)    & HR$^a$           & MR$^a$            & &     
 JS$^b$       &    JJL$^c$       &    \\ 
}
\startdata
J $=$ 0: &                  &                   & & 
                 &                  &       \\
1157.910A$^d$ & 26.97 $\pm$ 0.26 & $\ldots$     & & 
$\ldots$         & $\ldots$         & 28.09 \\
1157.910B$^d$ &  2.59 $\pm$ 0.30 & $\ldots$     & & 
$\ldots$         & $\ldots$         & 4.25 \\
1158.324A & 15.28 $\pm$ 0.23 & $\ldots$         & & 
$\ldots$         & $\ldots$         & 15.52 \\
1158.324B &  2.99 $\pm$ 0.35 & $\ldots$         & & 
$\ldots$         & $\ldots$         & 1.52 \\
1260.736A & $\ldots$         & 35.44 $\pm$ 0.33 & & 
37.0$^e$ $\pm$ 2.0 & 38.0$^e$ $\pm$ 2.6 & 31.27 \\
1260.736B & $\ldots$         &  3.17 $\pm$ 0.24 & & 
$\ldots$         & $\ldots$         & 4.88 \\
1328.833  & $\ldots$         & 48.23 $\pm$ 0.32 & & 
52.0$^e$ $\pm$ 4.0 & 51.0$^e$ $\pm$ 2.6 & 45.66 \\
\enddata
\tablenotetext{a}{ HR: High Resolution spectra, MR: Medium 
Resolution spectra.}
\tablenotetext{b}{ Jenkins \& Shaya 1979.}
\tablenotetext{c}{ Jenkins et al. 1983.}
\tablenotetext{d}{ A: Main component B: Secondary component, 
if resolved.}
\tablenotetext{e}{ Involves both the main and secondary $J=0$ 
components.}
\tablenotetext{f}{ Blended with $\lambda$1158.132 ($J$ $=$ 2).}
\tablenotetext{g}{ Total $W_{\lambda}$ for $\lambda\lambda$1329.085, 
1329.100, 1329.123.}
\tablenotetext{h}{ Total $W_{\lambda}$ for $\lambda\lambda$1261.426, 
1261.552.}
\tablenotetext{i}{ Total $W_{\lambda}$ for $\lambda\lambda$1329.578, 
1329.601.}
\end{deluxetable}

\begin{deluxetable}{cccccc}
\tablecolumns{6}
\tabletypesize{\footnotesize}
\tablewidth{0pc} 
\tablecaption{$W_{\lambda}$ (m\AA) of \ion{C}{1} Lines toward 
HD~112244}
\tablehead{
$\lambda$ & \multicolumn{2}{c}{Measured} & & 
\multicolumn{2}{c}{Calculated} \\ \cline{2-3} \cline{5-6}
 (\AA)    &  Component A  &  Component B  & &  
 Component A  &  Component B \\
}
\startdata
J $=$ 0:  &                &                 & &
                &                \\
1188.833$^a$  & 53.66 $\pm$ 2.42 & 32.44 $\pm$ 2.05
 & & 29.26 & 25.16 \\
1193.031$^b$  & 52.65 $\pm$ 2.46 & 47.22 $\pm$ 2.51
 & & 45.28 & 39.42 \\
1193.996  & 22.36 $\pm$ 1.91 & 25.10 $\pm$ 2.24 & &
 24.77 & 21.23 \\
1260.736  & 39.11 $\pm$ 1.20 & 27.60 $\pm$ 1.26 & &
 48.07 & 41.85 \\
1276.483  & 10.16 $\pm$ 0.68 & 14.27 $\pm$ 0.98 & &
 12.31 & 10.46 \\ 
1277.245$^c$ & 75.27 $\pm$ 1.28 & 63.47 $\pm$ 1.16 
& & 61.93 & 54.31 \\
1280.135  & 44.19 $\pm$ 1.40 & 29.04 $\pm$ 1.07 & & 
40.22 & 34.80 \\
\enddata
\tablenotetext{a}{ Neutral carbon lines are obscured by a strong 
Cl~{\small I}.}
\tablenotetext{b}{ Blended lines. The total $W_{\lambda}$ is listed 
for the stronger line.}
\tablenotetext{c}{ Blended lines. The total $W_{\lambda}$ is listed 
for the stronger line.}
\tablenotetext{d}{ Blended lines. The total $W_{\lambda}$ is listed 
for the stronger line.}
\tablenotetext{e}{ Blended lines. The total $W_{\lambda}$ is listed 
for the stronger line.}
\end{deluxetable}

\begin{deluxetable}{ccccccc}
\tablecolumns{7}
\tabletypesize{\footnotesize}
\tablewidth{0pc} 
\tablecaption{$W_{\lambda}$ (m\AA) of \ion{C}{1} Lines toward 
1~Sco}
\tablehead{
$\lambda$ & \multicolumn{2}{c}{Measured} & & 
\multicolumn{2}{c}{Other measurement} & Calculated \\
 \cline{2-3} \cline{5-6}
 (\AA)    & HR$^a$           & MR$^a$           & &
       \multicolumn{2}{c}{JJL$^b$}     
       &     \\
}
\startdata
J $=$ 0: &                  &                  & &  
                & &       \\
1193.031A$^c$ & 10.43 $\pm$ 0.62 & $\ldots$         
& & \multicolumn{2}{c}{$\ldots$}         &  9.35 \\
1193.031B & 11.46 $\pm$ 0.62 & $\ldots$         & & 
\multicolumn{2}{c}{$\ldots$}         & 9.56 \\
1193.996A$^d$ &  3.63 $\pm$ 0.42 & $\ldots$         
& & \multicolumn{2}{c}{$\ldots$}        &  3.62 \\
1193.996B$^d$ &  3.02 $\pm$ 0.62 & $\ldots$         
& & \multicolumn{2}{c}{$\ldots$}         & 3.37 \\
1260.736 & $\ldots$         & 19.73$^e$ $\pm$ 0.98 
& & \multicolumn{2}{c}{22.0$^e$ $\pm$ 2.4} & 20.18$^e$ \\
1277.245 & $\ldots$         & 69.20$^f$ $\pm$ 1.42 & & 
\multicolumn{2}{c}{$\ldots$}         & 37.48$^e$ \\
1280.135A & $\ldots$         &  7.38 $\pm$ 0.70 & & 
\multicolumn{2}{c}{$\ldots$}         &  7.11 \\
\enddata
\tablenotetext{a}{ HR: High Resolution spectra, MR: Medium 
Resolution spectra.}
\tablenotetext{b}{ Jenkins et al. 1983.}
\tablenotetext{c}{ Blended lines.}
\tablenotetext{d}{ A: Main component B: Secondary component, 
if resolved.}
\tablenotetext{e}{ Involves both the main and secondary $J=0$ 
components.}
\tablenotetext{f}{ Involves a second component and 
$\lambda$1277.282.}
\tablenotetext{g}{ Blended lines.}
\tablenotetext{h}{ Total $W_{\lambda}$ for $\lambda\lambda$1329.085, 
1329.100, 1329.123.}
\tablenotetext{i}{ Total $W_{\lambda}$ for $\lambda\lambda$1560.682, 
1560.709.}
\tablenotetext{j}{ Total $W_{\lambda}$ for $\lambda\lambda$1329.578, 
1329.601.}
\end{deluxetable}

\begin{deluxetable}{cccccc}
\tablecolumns{6}
\tabletypesize{\footnotesize}
\tablewidth{0pc} 
\tablecaption{$W_{\lambda}$ (m\AA) of \ion{C}{1} Lines toward 
$\pi$~Sco}
\tablehead{
$\lambda$ & Measured     & & \multicolumn{2}{c}{Other 
Measurement} & Calculated \\ \cline{4-5}
 (\AA)    &    & & \multicolumn{2}{c}{JS$^a$} 
 &             \\
}
\startdata
J $=$ 0: &                & &                  & &     
           \\
1260.736 &  3.40 $\pm$ 0.95 & &  \multicolumn{2}{c}{4.9 
$\pm$ 2.0} &  4.84    \\
1277.245$^b$ & 13.46 $\pm$ 0.49 & &  
\multicolumn{2}{c}{$\ldots$}    & 11.28  \\
1280.135 &  2.80 $\pm$ 0.50 & &  \multicolumn{2}{c}{$\ldots$}
        &  3.14  \\
1328.833 &  8.97 $\pm$ 0.32 & &  \multicolumn{2}{c}{$\ldots$}
        &  7.69  \\
1560.309 & 10.63 $\pm$ 0.66 & &  \multicolumn{2}{c}{$\ldots$}
        & 13.97  \\
1656.928 & 26.74 $\pm$ 0.74 & &  \multicolumn{2}{c}{$\ldots$}
        & 24.74  \\
J $=$ 1: &                  & &                  & &        
        \\
\enddata
\tablenotetext{a}{ Jenkins \& Shaya 1979.}
\tablenotetext{b}{ Involves $\lambda$1277.282.}
\tablenotetext{c}{ Total $W_{\lambda}$ for $\lambda\lambda$1329.085, 1329.100,
 1329.123.}
\tablenotetext{d}{ Total $W_{\lambda}$ for $\lambda\lambda$1560.682, 1560.709.}
\end{deluxetable}

\begin{deluxetable}{ccccccc}
\tablecolumns{7}
\tabletypesize{\footnotesize}
\tablewidth{0pc} 
\tablecaption{$W_{\lambda}$ (m\AA) of \ion{C}{1} Lines toward 
$\delta$~Sco}
\tablehead{
$\lambda$ & \multicolumn{2}{c}{Measured} & & 
\multicolumn{2}{c}{Other measurement} & Calculated 
\\ \cline{2-3} \cline{5-6}
 (\AA)    & HR$^a$           & MR$^a$           & &      
 \multicolumn{2}{c}{JS$^b$}     &      \\
}
\startdata
J $=$ 0: &                  &                  & &             
     & &       \\
1157.910A$^c$ & 23.06 $\pm$ 0.28 & $\ldots$         & & 
\multicolumn{2}{c}{$\ldots$}         & 20.36 \\
1157.910B$^c$ &  2.11 $\pm$ 0.20 & $\ldots$         & & 
\multicolumn{2}{c}{$\ldots$}         & 4.87 \\
1158.324A & 10.84 $\pm$ 0.23 & $\ldots$         & & 
\multicolumn{2}{c}{$\ldots$}         & 10.78 \\
1158.324B &  1.44 $\pm$ 0.25 & $\ldots$         & & 
\multicolumn{2}{c}{$\ldots$}         & 1.89 \\
1260.736A & 21.46 $\pm$ 0.43 & 28.53$^d$ $\pm$ 0.22 & & 
\multicolumn{2}{c}{32.0$^d$ $\pm$ 3.0} & 22.72 \\
1260.736B &  6.03 $\pm$ 0.36 & $\ldots$             & & 
\multicolumn{2}{c}{$\ldots$}         & 5.56 \\
J $=$ 1: &                  &                  & &     
             & &       \\
\enddata
\tablenotetext{a}{ HR: High Resolution spectra, MR: Medium Resolution 
spectra.}
\tablenotetext{b}{ Jenkins \& Shaya 1979.}
\tablenotetext{c}{ A: Main component B: Secondary component, if resolved.}
\tablenotetext{d}{ Involves both the main and secondary $J=0$ components.}
\tablenotetext{e}{ Blended lines.}
\tablenotetext{f}{ Blended lines.}
\end{deluxetable}

\begin{deluxetable}{ccccccc}
\tablecolumns{7}
\tabletypesize{\footnotesize}
\tablewidth{0pc} 
\tablecaption{$W_{\lambda}$ (m\AA) of \ion{C}{1} Lines toward 
$\omega^1$~Sco}
\tablehead{
$\lambda$ & \multicolumn{2}{c}{Measured} & & 
\multicolumn{2}{c}{Other measurements} & Calculated \\ 
\cline{2-3} \cline{5-6}
 (\AA)    &    $HST$         &  $Copernicus$       & & 
 SJ$^a$         &   JJL$^b$      &      \\
}
\startdata
J $=$ 0: &                  &                     & &    
            &                &       \\
1260.736 & 35.54 $\pm$ 0.14 &  $\ldots$           & & 26.0 
$\pm$ 1.7 & 33.0 $\pm$ 3.6 & 33.16 \\
1276.483 & $\ldots$         &  6.6 $\pm$ 4.1     & &  
$\ldots$      &  $\leq$ 6.4    &  8.85 \\
1328.833 & $\ldots$         & 22.2 $\pm$ 4.1     & &  
$\ldots$      & 62.0 $\pm$ 6.2 & 39.63 \\
J $=$ 1: &                  &                     & &      
          &                &       \\
1260.927 & 8.19 $\pm$ 0.12  &  $\ldots$           & &  4.7 
$\pm$ 1.6 &   $\leq$ 5.8   &  6.46 \\
1260.996 & 6.42 $\pm$ 0.11  &  $\ldots$           & &  2.7 
$\pm$ 1.6 &   $\leq$ 5.7   &  5.31 \\
1261.122 & 8.22 $\pm$ 0.13  &  $\ldots$           & & 11.0 
$\pm$ 1.7 &   $\leq$ 5.7   &  7.41 \\
\enddata
\tablenotetext{a}{ Snow \& Jenkins 1980.}
\tablenotetext{b}{ Jenkins et al. 1983.}
\tablenotetext{c}{ Total $W_{\lambda}$ for $\lambda\lambda$1329.085, 
1329.100, 1329.123.}
\tablenotetext{d}{ Total $W_{\lambda}$ for $\lambda\lambda$1329.578, 
1329.601.}
\end{deluxetable}

\begin{deluxetable}{ccccccc}
\tablecolumns{7}
\tabletypesize{\footnotesize}
\tablewidth{0pc} 
\tablecaption{$W_{\lambda}$ (m\AA) of \ion{C}{1} Lines toward 
$\sigma$~Sco \label{table:App2}}
\tablehead{
$\lambda$ & \multicolumn{2}{c}{Measured} & & 
\multicolumn{2}{c}{Other measurement} & Calculated \\ 
\cline{2-3} \cline{5-6}
 (\AA)    & HR$^a$           & MR$^a$           & &      
 \multicolumn{2}{c}{JS$^b$}     &      \\
}
\startdata
J $=$ 0: &                  &                  & &       
             &                &       \\
1260.736A$^c$ & 23.59 $\pm$ 0.64 & 36.08$^d$ $\pm$ 0.50 
& & \multicolumn{2}{c}{23.0 $\pm$ 2.0} & 26.59 \\
1260.736B$^c$ &  2.41 $\pm$ 0.41 & $\ldots$         & & 
\multicolumn{2}{c}{$\ldots$}         & 4.40 \\
1328.833 & $\ldots$         & 50.90$^d$ $\pm$ 0.50 & & 
\multicolumn{2}{c}{$\ldots$}         & 39.57 \\
1560.309 & $\ldots$         & 56.50$^d$ $\pm$ 0.54 & & 
\multicolumn{2}{c}{$\ldots$}         & 55.97 \\
J $=$ 1: &                  &                  & &     
               &                   &       \\
1260.927 & 12.46 $\pm$ 0.51 &  9.40 $\pm$ 0.50 & & 
\multicolumn{2}{c}{$\ldots$}         & 11.90 \\
1260.996 & 10.04 $\pm$ 0.47 & 11.10 $\pm$ 0.50 & & 
\multicolumn{2}{c}{$\ldots$}         &  9.83 \\
\enddata
\tablenotetext{a}{ HR: High Resolution spectra, MR: Medium 
Resolution spectra.}
\tablenotetext{b}{ Jenkins \& Shaya 1979.}
\tablenotetext{c}{ A: Main component B: Secondary component, 
if resolved.}
\tablenotetext{d}{ Involves both the main and secondary $J=0$ 
components.}
\tablenotetext{e}{ Total $W_{\lambda}$ for $\lambda\lambda$1329.085, 
1329.100, 1329.123.} 
\end{deluxetable}

\clearpage
\addtocounter{figure}{-1}

\clearpage
\begin{figure}[htbp] 
\plotone{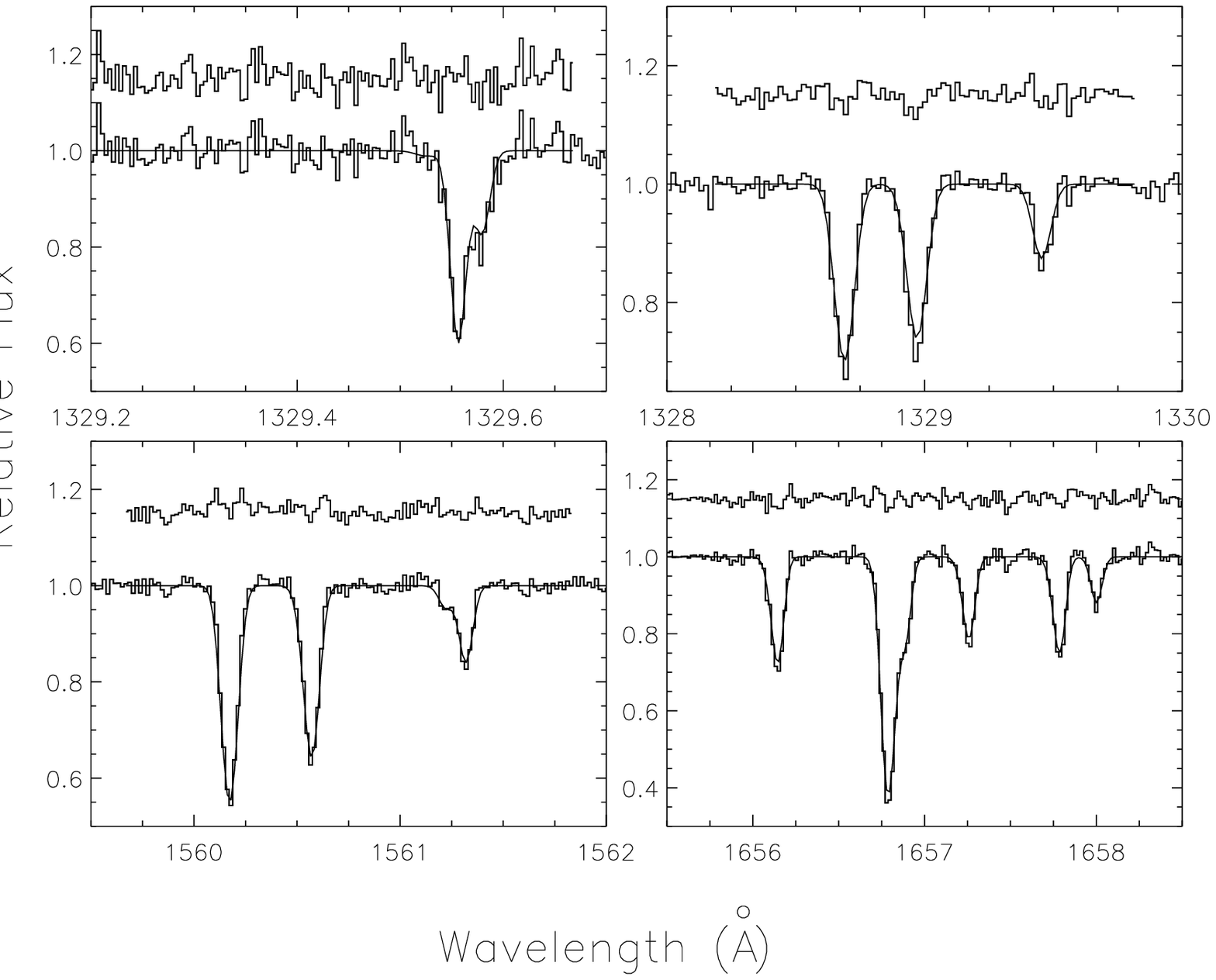}
\caption{b) Spectra of multiplets $\lambda$$\lambda$1329 (high 
resolution, $J$ $=$ 2 level only), 1329, 1561, and 1657.}
\end{figure}

\clearpage
\begin{figure}[htbp] 
\epsscale{0.65}\plotone{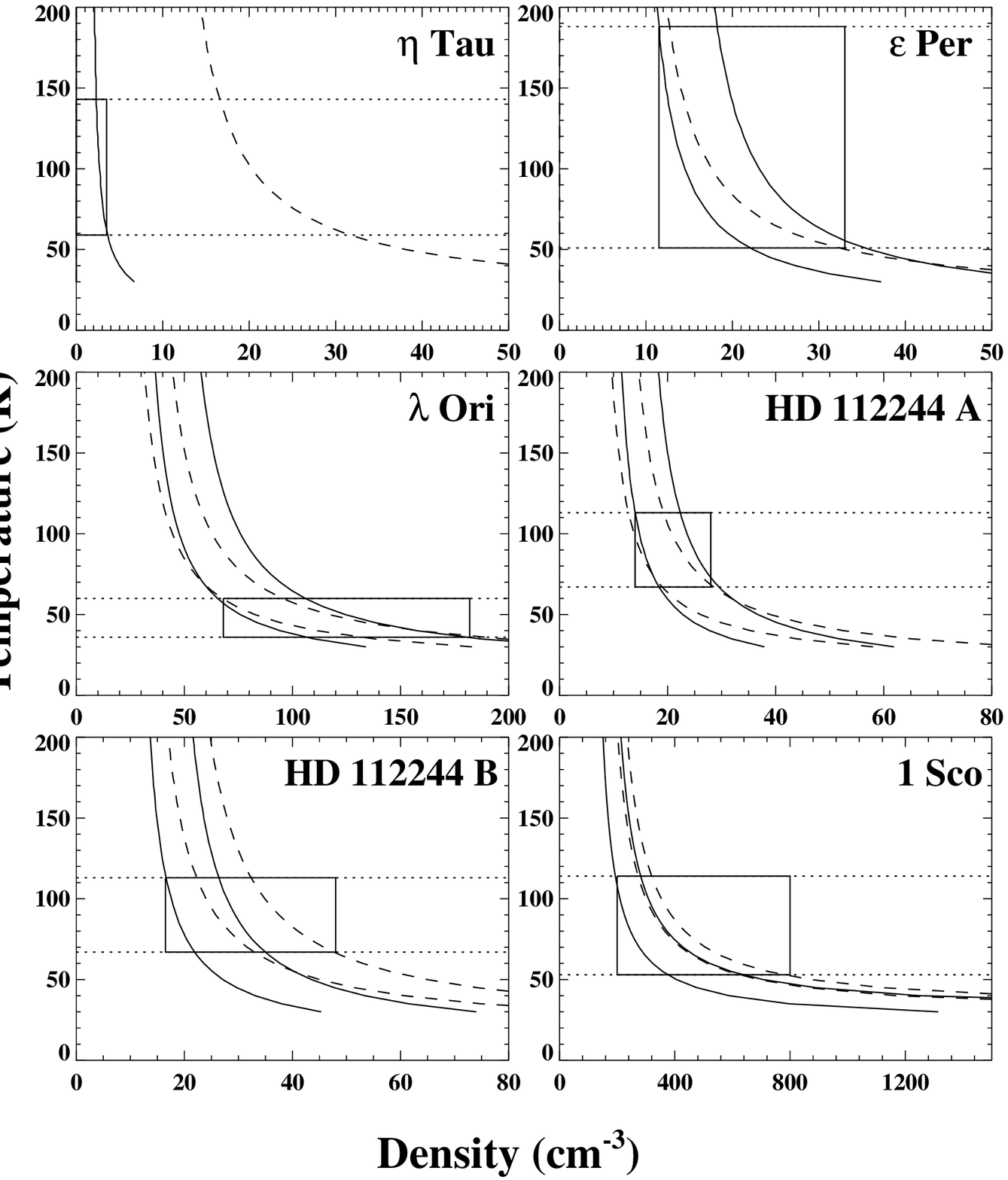}
\caption{The density-temperature relationships for the 
main components along our sight
lines. The curves for the secondary component toward HD~112244 
(component B) are also displayed. 
The two solid lines represent the density-temperature correspondence 
calculated from the lowest and 
the highest $n_{J=1}/n_{J=0}$ ratios allowed by our \ion{C}{1} 
column densities. A similar pair of lines  
(dashed lines) show the relationships calculated from 
the $n_{J=2}/n_{J=0}$ ratios.
If only an upper limit was available for either the $J$ $=$ 1 or 2 
column density, then a single 
line shows the maximum gas densities as a function of temperature 
for the corresponding ratio.  The density--temperature values, inferred 
from a ratio, are between the curves belonging to the same ratio.
In the cases of upper limits, all density-temperature points that are 
to the left of the curve are allowed.
The dotted horizontal lines are the H$_2$ excitation temperatures 
from Savage et al. 1977.  The thick rectangles 
contain all density--temperature values that are consistent with
both ratios and the H$_2$ excitation temperatures. The overlap
between the ranges was not sufficient toward HD~112244~B, 1~Sco, and
$\beta^1$~Sco; in these cases, the rectangles contain the extreme 
density--temperature values that are consistent
with all measures. \label{DTcurves}}
\end{figure}
\addtocounter{figure}{-1}

\clearpage
\begin{figure}[htbp] 
\epsscale{0.65}\plotone{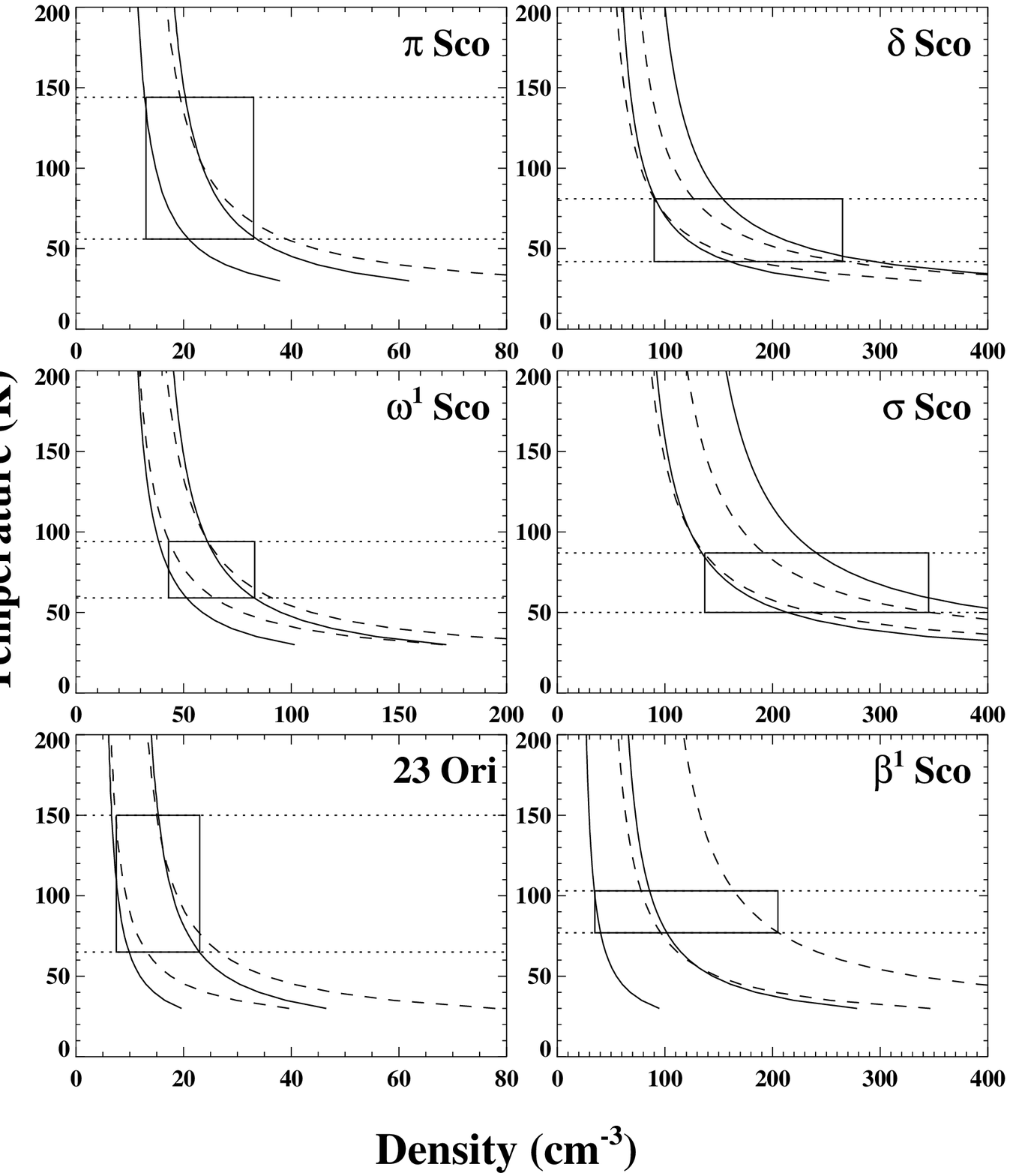}
\caption{Cont.}
\end{figure}

\clearpage
\begin{figure}[htbp] 
\plotone{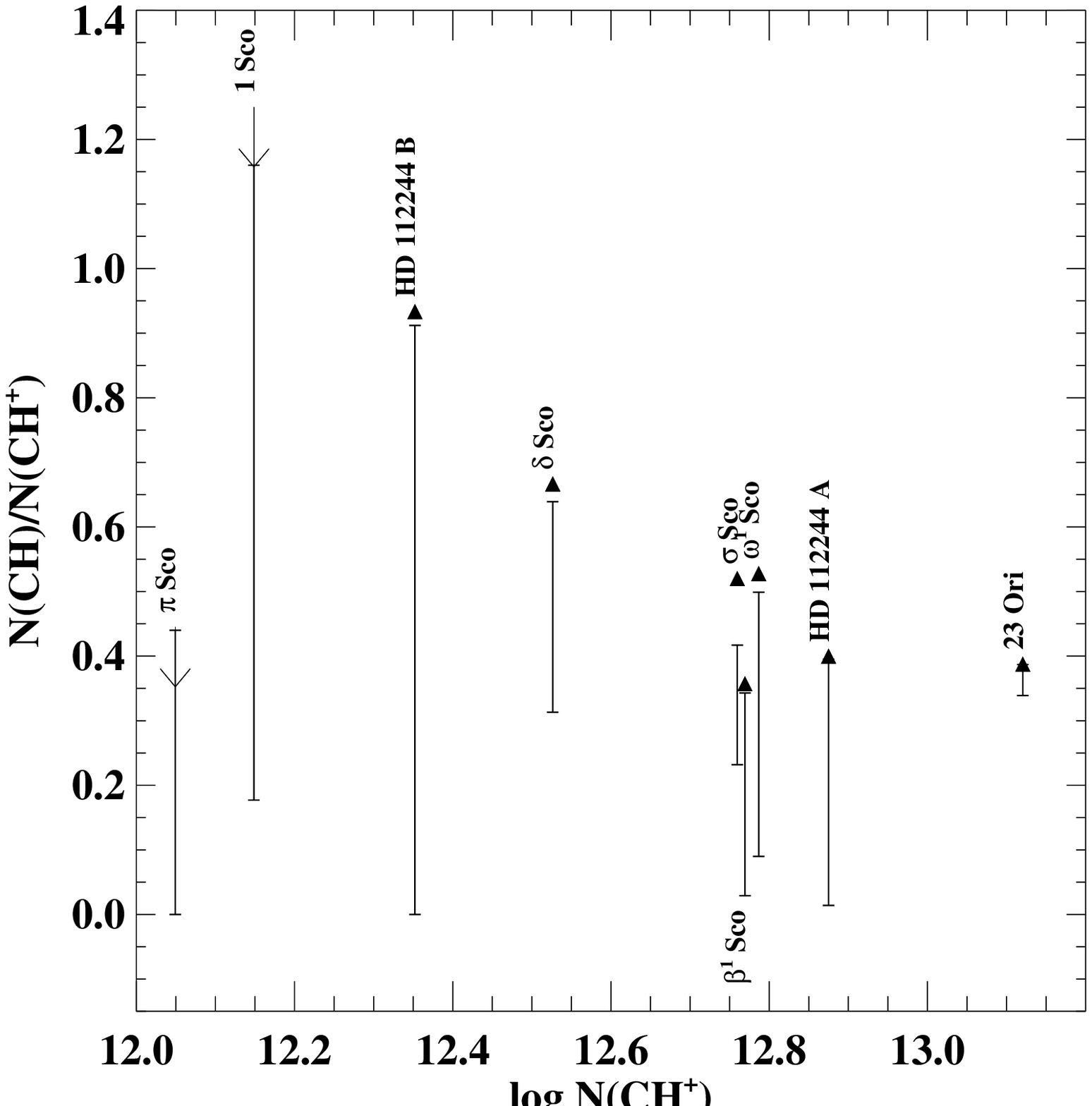}
\caption{CH produced by the non-equilibrium CH$^+$ 
chemistry, normalized by the 
observed amount of CH$^+$, as a logarithmic function of CH$^+$ 
columns. The arrows and triangles indicate
the upper limits and measured amount of CH, respectively. The maximum and 
the minimum amount of CH 
associated with the CH$^+$ chemistry are shown by ranges. The identifier of 
each sight line is printed 
above the symbols except for $\beta^1$~Sco which
is below the symbol for clarity. \label{thermal_chem}}
\end{figure}

\end{document}